\documentstyle[11pt,aaspp4]{article}

\def\etal{{\it et~al.\/\ }}

\def\teff{{\it T$_{\rm eff}$}}
\def\logg{{\rm log~g}}

\def\kms{{km~s$^{-1}$}}

\def\Rm31{{R$_{\rm M31}$}}

\def\hii{{\ion{H}{2}\/\ }}

\def\R23{{R$_{23}$}}

\slugcomment{In preparation for the Astrophysical Journal}

\lefthead{Venn et al.}
\righthead{Stellar Abundances in NGC\,6822}

\begin{document}

\title
{First Stellar Abundances in NGC\,6822 from VLT-UVES
and Keck-HIRES Spectroscopy
	\footnote{Based on observations obtained from the UVES
	commissioning at the Very Large Telescope (Kueyen),
	European Southern Observatory, Paranal, Chile, 
	and the W.\,M.\,Keck Observatory, Hawaii.  
	Keck is operated as a scientific partnership 
	among the California Institute of Technology, the University of
	California, and the National Aeronautics and 
	Space Administration, and was made possible by the 
	generous financial support of the W.\,M.\,Keck Foundation.} }

\author{K. A. Venn\altaffilmark{2} }
\affil{Macalester College, Saint Paul, MN, 55105  
(venn@clare.physics.macalester.edu)}

\author{D. J. Lennon} 
\affil{Isaac Newton Group of Telescopes (ING), 
Santa Cruz de La Palma, Canary Islands, E-38780, Spain
(djl@ing.iac.es)}

\author{A. Kaufer}
\affil{European Southern Observatory, Alonso de Cordova 3107, Santiago, 
Chile (akaufer@eso.org)}

\author{J. K. McCarthy\altaffilmark{3} }
\affil{Palomar Observatory, California Institute of Technology, 
Pasadena, CA, 91125 (jkmccarthy@pacbell.net)}

\author{N. Przybilla and  R.\,P. Kudritzki\altaffilmark{4}}
\affil{Institut f\"ur Astronomie und Astrophysik, Universit\"ats-Sternwarte
M\"unchen, M\"unchen, D-81679, Germany (nob@usm.uni-muenchen.de and
kudritzki@usm.uni-muenchen.de)}

\author{M. Lemke\altaffilmark{5}}
\affil{Dr.\,Karl Remeis-Sternwarte, Bamberg, D-96049, Germany
(ai26@sternwarte.uni-erlangen.de)}

\author{E.D. Skillman}
\affil{Department of Astronomy, University of Minnesota, 
Minneapolis, MN, 55455 (skillman@astro.umn.edu)} 

\and

\author{S. J. Smartt} 
\affil{Institute of Astronomy, University of Cambridge,
Madingley Road, Cambridge, CB3 0HA (sjs@ast.cam.ac.uk)}

\altaffiltext{2}{Adjunct Assistant Professor, Department of Astronomy,
University of Minnesota, Minneapolis, MN, 55455 } 
\altaffiltext{3}{Present address:
Pixel Vision, Inc., Advanced Imaging Sensors Division, 
4952 Warner Avenue, Suite 300, Huntington Beach, CA, 92649. 
Also, Adjunct Assistant Professor of Physics, Loyola Marymount 
University, 7900 Loyola Blvd., Los Angeles, CA, 90045.}   
\altaffiltext{4}{Max Planck Institut f\"ur Astrophysik, 
Garching, D-85740, Germany} 
\altaffiltext{5}{Present Address: INA-Werk Schaeffler, Herzogenaurach, Germany}

\begin{abstract}

We have obtained the first high resolution spectra of individual
stars in the dwarf irregular galaxy, NGC\,6822.   The spectra of
the two A-type supergiants were obtained at the VLT and Keck 
Observatories, using UVES and HIRES, respectively.   A detailed
model atmospheres analysis has been used to determine their
atmospheric parameters and elemental abundances.   The mean
iron abundance from these two stars is 
$<$[Fe/H]$>=-$0.49 $\pm$0.22 ({\it $\pm$0.21})\footnote{In this paper, 
abundances shall be reported with two uncertainties; the first is the 
line-to-line scatter, and the second (in parentheses and italics) is an 
{\it estimate} of the systematic error due to uncertainties in the 
atmospheric parameters.}, 
with Cr yielding a similar 
underabundance, $<$[Cr/H]$>=-$0.50 $\pm$0.20 ({\it $\pm$0.16}).  
This confirms that NGC\,6822 has a metallicity that is slightly 
higher than that of the SMC, and is the
first determination of the present-day iron-group abundances 
in NGC\,6822.
The mean stellar oxygen abundance,
12+log(O/H)=8.36 $\pm$0.19 ({\it $\pm$0.21}),
is in good agreement with the nebular oxygen results.
Oxygen has the same underabundance as iron, 
$<$[O/Fe]$>$=+0.02 $\pm$0.20 ({\it $\pm$0.21}). 
This O/Fe ratio is very similar to that seen in the 
Magellanic Clouds, which supports the picture that chemical 
evolution occurs more slowly in these lower mass galaxies,
although the O/Fe ratio is also consistent with that
observed in comparatively metal-poor stars in the Galactic disk.
Combining all of the available abundance observations for
NGC\,6822 shows that there is no trend in abundance with
galactocentric distance.   However, a subset of the highest
quality data are consistent with a radial abundance gradient.
More high quality stellar and nebular observations are 
needed to confirm this intriguing possibility.

\end{abstract}

\keywords{stars: abundances, atmospheres, supergiants --- 
galaxies: abundances, individual (NGC6822), stellar content}

\section{Introduction}

Understanding the evolution of chemical abundances in galaxies provides
an important constraint for uniquely determining their star formation histories.
This is because prior stellar nucleosynthesis information is preserved 
in the metallicity distribution of both the stars and gas.    
The analysis of bright nebular emission 
lines of \hii regions has been the most frequent approach to modelling 
chemical evolution of galaxies to date (c.f., Pagel 1997).
And yet, only a very limited number of elements can be examined and
quantified when using this approach.   The chemical evolution of
a galaxy depends on the contributions of all its ISM-enriching 
constituents (e.g., type I supernovae, high mass stars, thermal pulsing 
intermediate AGB stars).  Thus, more elements than just 
those observed from nebular studies need to be measured, since each has 
a different formation site which sample different constituents
(e.g., oxygen is created primarily in massive stars, while iron comes
from supernovae of both high and low mass stars).
The need for abundances of heavy elements, and the new opportunities 
made possible by the 8- to 10-m telescopes and efficient high resolution
spectrographs,
have motivated us to determine elemental abundances in young stars
in nearby galaxies.   In this paper, we present our first results
from an analysis of two A-type supergiants in NGC\,6822.

NGC\,6822 is a well-studied Local Group dwarf irregular galaxy
with a distance modulus of 23.49 $\pm$0.08 (Gallart \etal 1996a).  
Its stellar content has been extensively examined 
(see Chapter~9 in van den Bergh 2000 for a recent review),
and it is known to have a low metallicity from nebular analyses. 
Its oxygen abundance was found to be 12+log(O/H)=8.25 $\pm$0.07 
from a study of seven \hii regions (Pagel \etal 1980, or
8.14 $\pm$0.08 considering only those \hii regions where 
\ion{O}{3}~$\lambda$4363 was detected).   These results are also supported by
oxygen abundances from planetary nebulae (Richer \& McCall 1995,
Dufour \& Talent 1980).
This result is slightly higher than in the SMC 
(12+log(O/H)=8.1), as found from both nebular and 
stellar analyses (c.f., Garnett \etal 1995, Venn 1999 
and references therein).

A metallicity intermediate between the SMC and the LMC 
has also been reported from 
indirect observational data, e.g., the colors and spectral type
distributions of red supergiants (Elias \& Frogel 1985)
and the relative color of an S star (Aaronson \etal 1985).
Also, the number of WR stars to the number of massive stars
in NGC\,6822, which appears to be related to the metallicity 
of the galaxy, is between the number ratios in the LMC and SMC
(Armandroff \& Massey 1985, Azzopardi \etal 1988). 
However, a study of OB stellar spectral types by Massey \etal (1995) 
suggested that NGC\,6822 may have a somewhat lower metallicity 
than the SMC based on weak metal line strengths. 
This claim has been refuted by a recent low resolution study of
blue supergiants using VLT FORS by Muschielok \etal (1999).
Muschielok \etal\ also suggest a mean metallicity of $-$0.5 $\pm$0.2\,dex
relative to the Sun, although their spectra do not sample the iron-group.

The stars in NGC\,6822 are interesting themselves as well.
Studies of very low-metallicity, massive stars need to be done 
in the dwarf irregular galaxies (the SMC being the most familiar) 
since extinction within the Galactic plane makes it too difficult 
to find and study any more nearby metal-poor young, massive stars 
in the outskirts of the Galaxy.  
Metallicity is recognized as a fundamental parameter in stars that
affects star formation, stellar evolution, and stellar wind parameters.
Thus, spectral analyses of stars in dwarf irregular galaxies offer a unique 
opportunity; in particular, the results can be compared to the wealth 
of information published on stars in the Galaxy and Magellanic Clouds.

\section{Observations and Reductions}

Two A-type supergiants were initially selected from the literature
(van den Bergh \& Humphreys 1979); coordinates and color information 
are listed in Table~\ref{basic}.   One of us (DJL) obtained low
resolution spectra using EFOSC on the ESO 3.6m telescope on 17 September 
1996, as part of a program to determine accurate spectral types for the 
brightest stars in Local Group galaxies.   
EFOSC was used in long-slit mode, with a 1.5'' slit aligned on both 
stars.   The B150 grism gave a resolution of $\sim$8\,\AA\ and wavelength 
coverage from 3780 to 5510\,\AA.  A single 30 minute exposure yielded 
S/N between 30 and 50, depending on wavelength.  These observations 
confirmed that both stars are early A-type supergiants in NGC\,6822,
and thus they became our top priority for follow-up high resolution 
observations. 

Observations of Star cc and Star m were made at Keck using HIRES
(Vogt \etal 1994)
on 26 \& 27 September 1997 and 6 \& 7 October 1999, respectively.  
Both stars received three one-hour exposures, and an extra 
half-hour exposure was made for Star cc.   All observations were made 
through less than optimal conditions, including high humidity for
Star cc and thin cirrus for Star m.  For Star cc, seeing ranged from 
1.0'' to 1.3'', thus a 1.1'' slit was used, yielding resolving power 
of R=35,000 over a 4 pixel resolution element.   For Star m, subarcsecond 
seeing permitted a 0.86'' slit, and thus slightly higher resolving power
(R=45,000).  The spectra span 4300 $\le$ $\lambda$ $\le$ 6700 \AA\ in 30 
echelle orders,
although the wavelength coverage is incomplete beyond $\lambda$5200
on the TK2048 CCD used.   Slit length was limited to 7.0''
to prevent overlapping orders at the short wavelength extreme.
The two-dimensional CCD echelle spectrograms were reduced by JKM 
using the Figaro package, and a set of routines written specifically
for echelle data reduction (c.f.  McCarthy \etal 1995, 
McCarthy \& Nemec 1997).  Contamination by night sky lines and 
emission nebulae were removed from the stellar spectra prior to extraction 
by fitting low-order polynomials to ``sky aperatures'' adjacent to the stellar 
spectrum.   Following optimal extraction (with cosmic ray removal), each
echelle order was rebinned from 2048 to 1024 channels to increase the S/N
per channel. 
The signal-to-noise in the rebinned and coadded HIRES spectra for Star cc 
is S/N = 35-56 per 2-channel resolution element.  The HIRES S/N for Star m
was lower, S/N = 21-45 per 2-channel resolution element.   In both cases, 
the S/N improves in redder orders and decreases away from the center of each 
order.

An improved spectrum of Star m was obtained by AK with three two-hour exposures 
using UVES (D'Odorico \& Kaper 2000) at the VLT on three consecutive nights, 
8-10 October 1999, as part of the first UVES commissioning run at the 
second unit telescope UT2 (``Kueyen'').
The spectrograph was set up in a predefined standard dichroic mode
(Dichroic\,\#1) with the blue central wavelength on cross-disperser \#2 set
to 390\,nm and the red cross-disperser \#3 to 564\,nm, which gives nearly
full wavelength coverage from $<$3600 to 6657 \AA\ (missing small blocks
from only 5610-5700\,\AA\ due to the dichroic mirror, and 4524-4637\,\AA\
due to the gap in between the red CCD mosaic). The three detectors in the
two arms were used without binning and in low gain giving a read-out
noise of $<4$\,electrons. The seeing conditions varied from 0.5'' to 0.9''
over the three exposures, with an average seeing of 0.7''. 
The spectrograph slit was set to a width of 1.0''. 
Star m is located in a crowded field with nearby faint companions 
($\sim$2-5'' separations).  The field was derotated and the slit kept
fixed at a constant position angle of 90 degrees. This further required
the use of the atmospheric dispersion compensator (ADC) to optimize the
slit throughput over the large wavelength range in the dichroic setting.
The raw spectra are publically available through the ESO archive and
were reduced with the ESO-MIDAS reduction package using the new UVES
context (which is also the basis of the fully automatic data-reduction
pipeline running at the telescope).  For Star m, an optimum extraction 
algorithm with simultaneous cosmic ray rejection and sky-background subtraction 
was used. The three single flat-fielded and wavelength calibrated spectra 
from each spectrograph arm were then simply averaged with weights according 
to their S/N.  The resultant spectra have resolution R=30,000 over a 
3 pixel resolution element.  A signal-to-noise ratio of S/N = 50-70 per 
resolution element, was attained after coaddition. 

\begin{figure}
\plotone{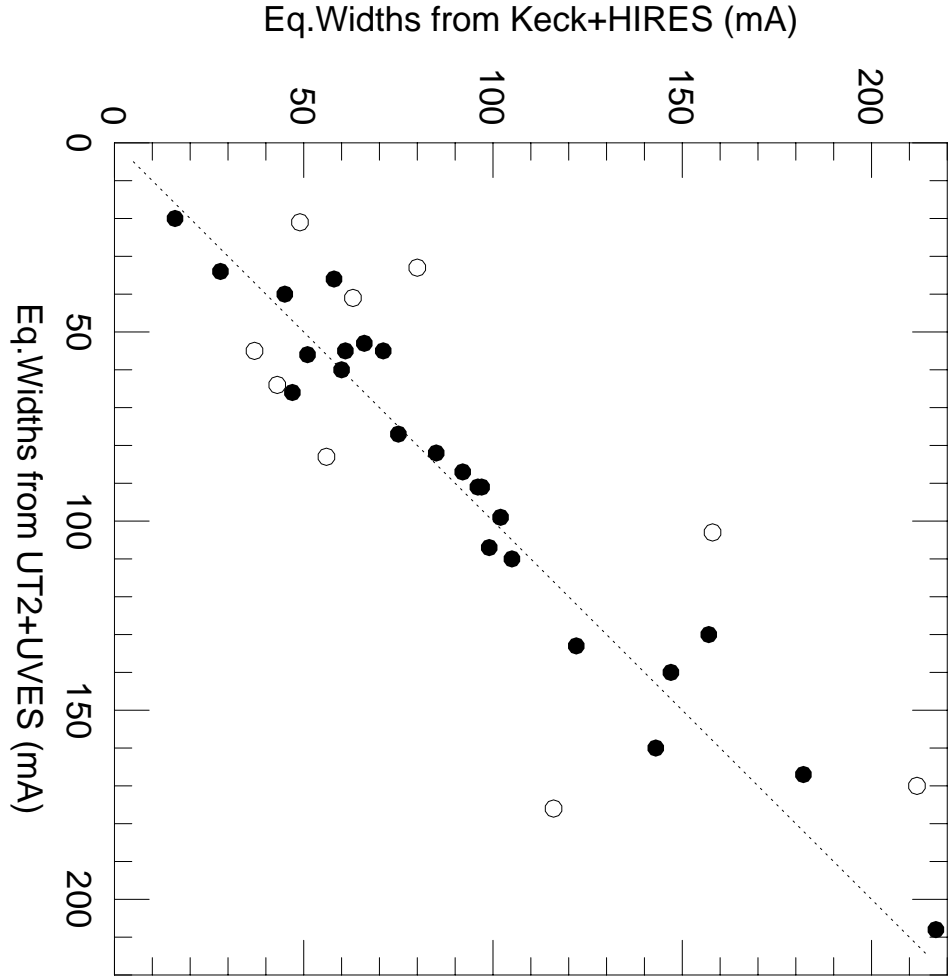}
\caption{A comparison of equivalent width measurements from 
a subset of lines in Star m from the Keck+HIRES spectrum versus 
the UT2+UVES spectrum ({\it hollow circles} are measurements of 
lines from the low S/N regions of the HIRES spectrum).
No significant offsets are seen.  Thus, given the differences in 
exposure times, observing conditions, and set-ups, we verify that 
the two instruments yield comparable spectra. 
\label{starm}}
\end{figure}

Table~\ref{lines1} lists our identifications and equivalent width
measurements for most of the stellar absorption lines in the spectra
of these two stars.  Very strong and/or blended lines have been neglected
(though some strong lines are noted in italics, and some blended lines
noted by ``b'', in cases where the line is listed because it is used 
in the other star).   
Figure~\ref{starm} shows a comparison of equivalent width measurements
for 33 lines in common between the UVES and the HIRES spectra for Star~m.   
No significant offsets are apparent, 
and given the differences in the exposure times, S/N, and different 
observing conditions (e.g., weather, difference in airmass to NGC\,6822),
then we verify that the two instruments yield comparable spectra. 
The overall quality of our UVES spectrum of Star m is better
(S/N and wavelength coverage), thus our analysis of Star m throughout
this paper, and the equivalent widths listed in Table~\ref{lines1}, 
are entirely from the UVES spectrum.  
Given the range in the S/N of the final spectra, 
the equivalent widths in Table~\ref{lines1} have an estimated 
uncertainty of $\sim$10\% (highest S/N) to $\sim$20\%.

\section{Atmospheric Analyses}

Both stars have been analysed using ATLAS9 (hydrostatic,
line-blanketed, plane parallel) model atmospheres (Kurucz 1979, 1988).
These atmospheres have been used successfully for photospheric 
analyses of A-F supergiants in the Galaxy, the Magellanic Clouds, 
and M31 (Venn \etal 2000, Venn 1999, 1995a, 1995b, Luck \etal 1998, 
Hill 1999, 1997, Hill \etal 1995). 

Analyses of A-type supergiants requires a tailored analysis,
where only weak spectral lines (preferrably that form deep in
the photosphere) are included.  
Weak lines are defined as lines where a change in microturbulence
($\xi$, discussed further below), 
$\Delta\xi=\pm$1\,\kms\ yields a change in abundance of 
log($X$/H)$\le\pm$0.15.   
Typically, W$_\lambda\le$200\,m\AA.
Using weak lines exclusively helps us to avoid uncertainties in the model
atmospheres analysis due to neglected non-LTE and spherical extension
effects in the atmospheric structure, as well as non-LTE and
$\xi$ effects in the line formation calculations. 
 
\begin{figure}
\plotone{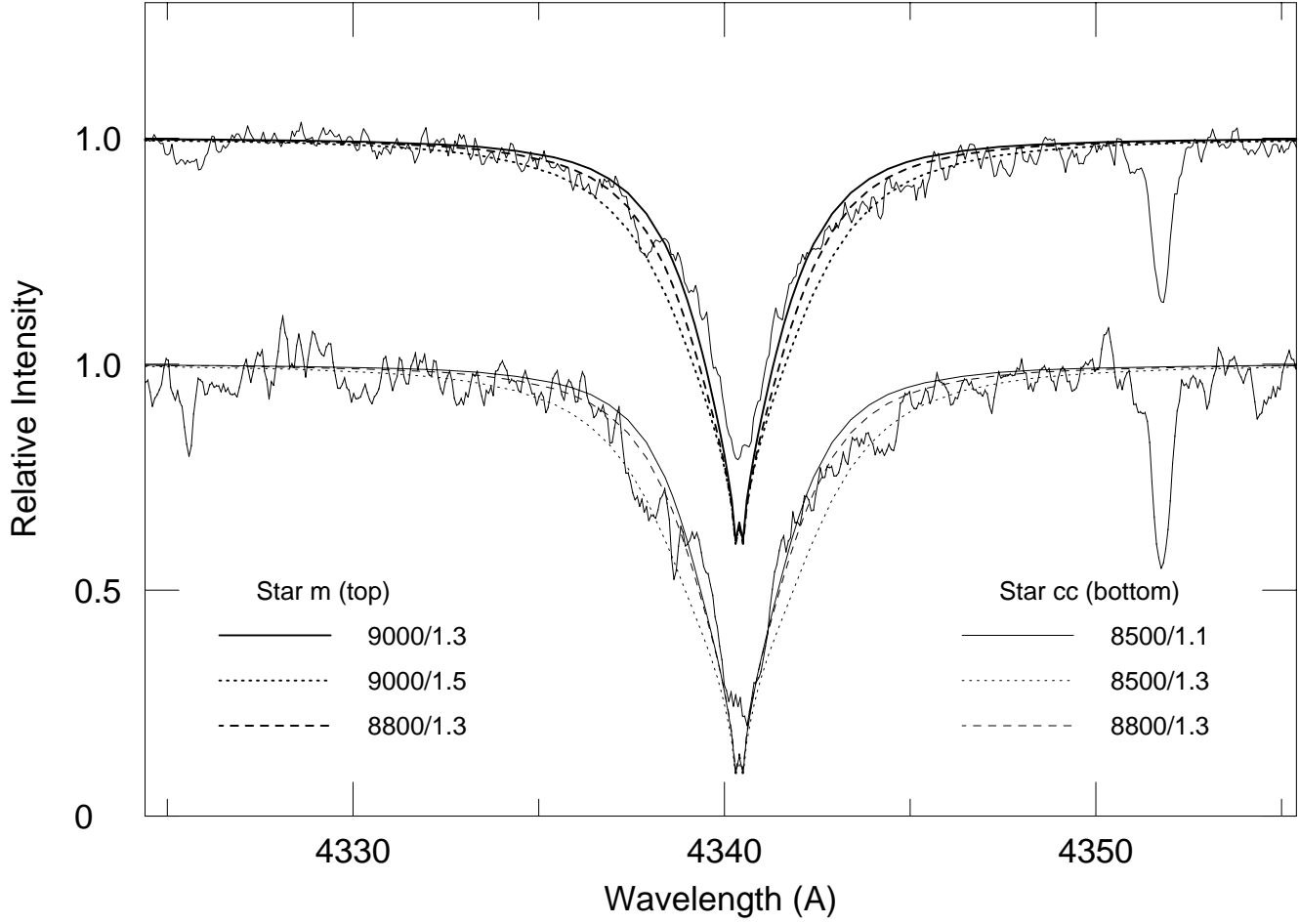}
\caption{H$\gamma$ profiles for Star m (top) and Star cc (bottom).  
Theoretical fits are shown for \teff/\logg\ pairs for both stars.
\label{hgamma}}
\end{figure}

\begin{figure}
\plotone{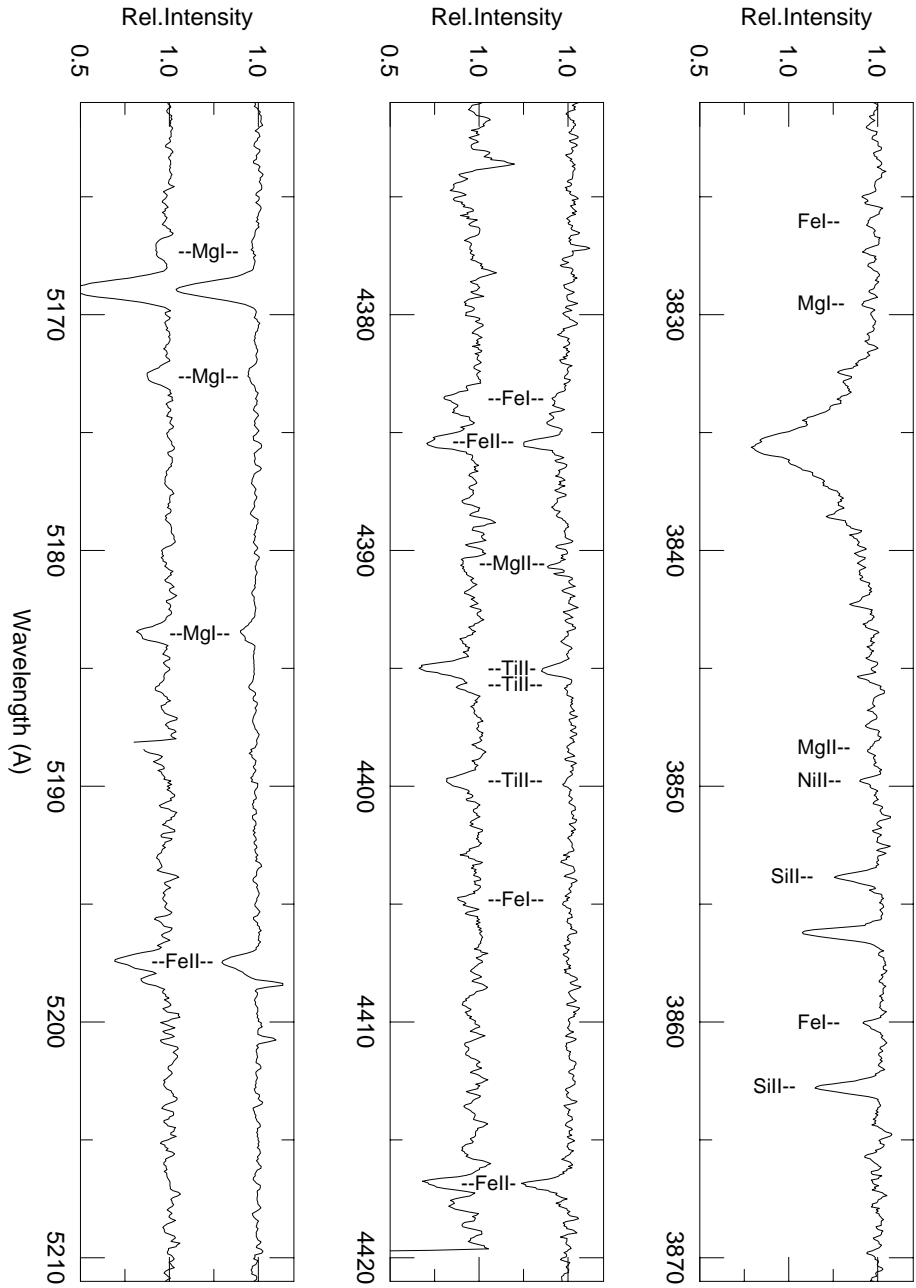}
\caption{Sample spectra of Star m (top) and Star cc (bottom)
around the critical \ion{Mg}{1} and \ion{Mg}{2} lines.
\label{spec-mg}}
\end{figure}

\begin{figure}
\plotone{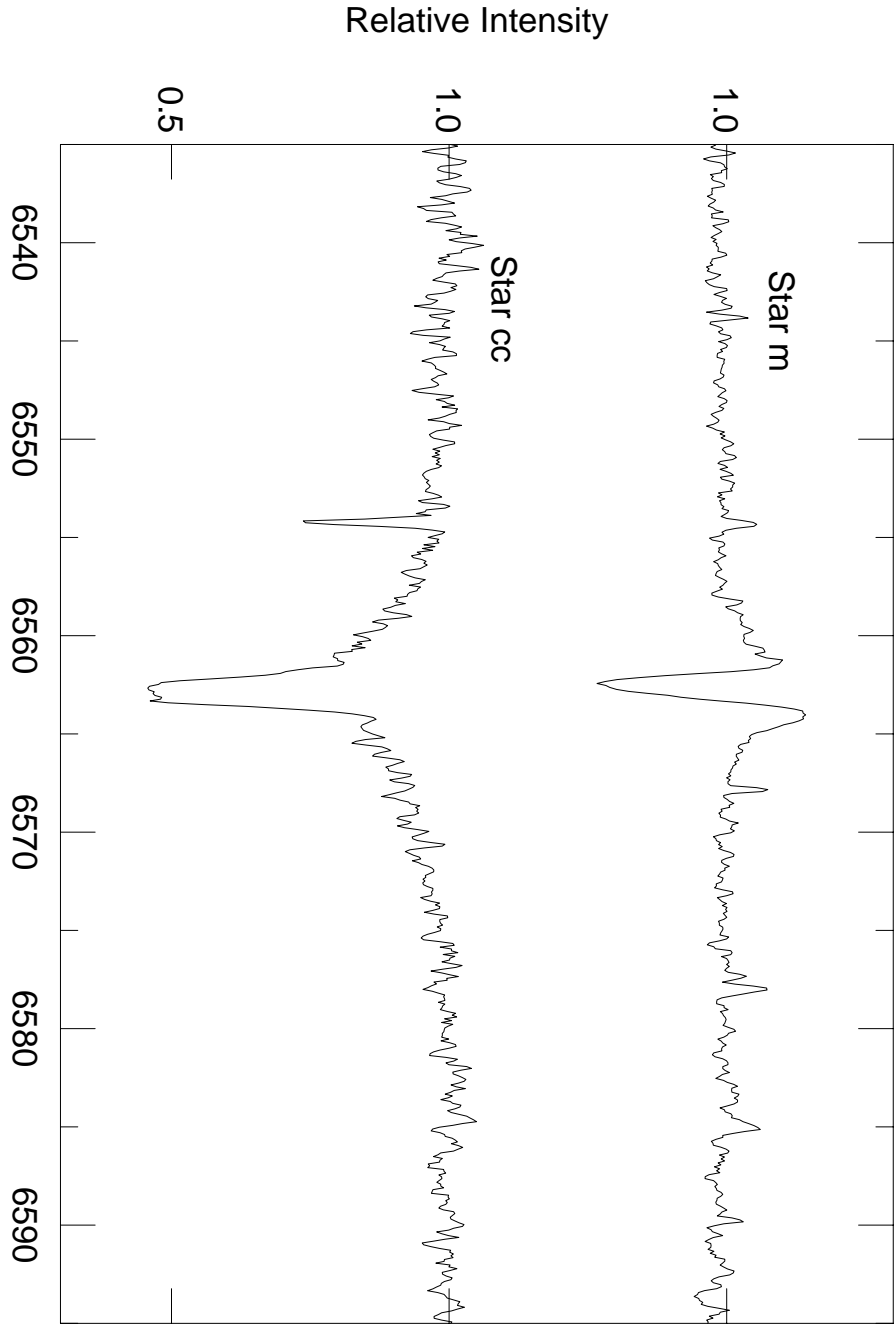}
\caption{H$\alpha$ spectra for Star m (top) and Star cc (bottom).
Neither shows evidence of a strong stellar wind, such as 
a P~Cygni profile.
\label{halpha}}
\end{figure}

\begin{figure}
\plotone{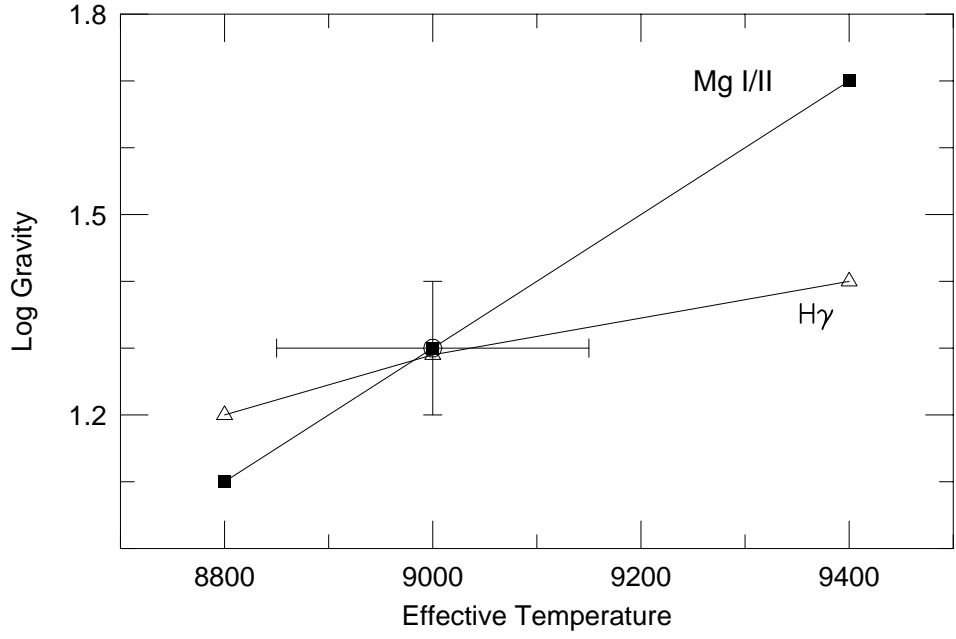}
\caption{\teff-gravity pairs that fit the spectral features
of Star m.   {\it Hollow triangles} track H$\gamma$ fits;  
{\it filled squares} follow the Mg ionization equilibrium.   
\label{atm-m}}
\end{figure}

The critical spectral features 
used to determine the model atmosphere parameters (effective 
temperature, \teff, and gravity) are the wings of the H$\gamma$ 
line (e.g., see Figure~\ref{hgamma}) and ionization equilibrium 
of \ion{Mg}{1} and \ion{Mg}{2} (e.g., see Figure~\ref{spec-mg}).   
H$\gamma$ does not appear to be affected by a stellar wind 
component in either star; in fact, examination of H$\alpha$ 
in both stars (see Figure~\ref{halpha}) shows that neither has 
a strong wind (although some wind signatures can still be seen
in both stars). 
A locus of \teff-gravity pairs that reproduce the observed
H$\gamma$ profile and another locus where Mg ionization 
equilibrium occurs were examined, and their intersection
point adopted for the best atmospheric parameters per star
(e.g., see Figure~\ref{atm-m}).

Non-LTE calculations are included for Mg
using the model atom developed by Gigas (1988) and a system of
programs first developed by W.\,Steenbock at Kiel University
and further developed and upgraded by M.\,Lemke.  
Mg non-LTE calculations in Galactic A-F supergiants have been described
by Venn (1995b) for all but the lines near 3850\,\AA. 
The non-LTE corrections for all Mg lines used in this analysis are 
typically small ($\le$0.23~dex), but including the corrections improves 
the atmospheric parameter determinations.   For the two stars
analysed here, the Mg non-LTE corrections are shown line by line
in Table~\ref{mgnlte}.

Comparing the two ionization states of iron cannot currently be used 
as a reliable atmospheric parameter indicator due to \ion{Fe}{1} 
non-LTE effects, which are more complicated to model than Mg.  
Estimates of the log(\ion{Fe}{1}/H) non-LTE corrections range from
$-$0.2 to $\ge$$-$0.3~dex (c.f., Boyarchuk \etal 1985, Gigas 1986).
However, non-LTE effects are expected to be negligible for
\ion{Fe}{2} lines, the dominant species of iron, as 
confirmed from detailed calculations by Becker (1998). 

The microturbulence was found by examining the line 
abundances of \ion{Fe}{2} and \ion{Cr}{2}, and requiring
no relationship with equivalent width.   Examination of
the \ion{Ti}{2} lines suggested a lower microturbulence
($\sim-$1 to $-$2~\kms) for this species.   Considering the 
weak lines nature of this analysis, and the estimated 
uncertainty of $\Delta\xi=\pm$1~\kms, a single value 
for $\xi$ was adopted for each star throughout.

The atmospheric parameters determined for both stars are listed
in Table~\ref{basic}.   Based on these parameters, new spectral
types are assigned (by comparison to Galactic A supergiants with
similar atmospheric parameters, see Venn 1995a).   
Uncertainties in \teff\ are estimated from the range where 
log(\ion{Mg}{2}) = log(\ion{Mg}{1})$\pm$0.2 when holding gravity 
fixed.  This range allows for uncertainties in equivalent width 
measurements, atomic data, and uncertainties in the non-LTE 
calculations.  Similarly, uncertainties in gravity are estimated 
from the range in the H$\gamma$ profile fits 
while holding \teff\ fixed (e.g., see Figure~\ref{hgamma}).

Intrinsic (B-V)$_o$ colors and bolometric corrections for each 
star are also determined from the ATLAS9 model atmospheres using 
the Kurucz program UBVBUSER\footnote{Program available 
from R.\,L.\,Kurucz at {\it http://cfaku5.harvard.edu/programs.html}}.
For Star~cc and Star~m, (B-V)$_o$=+0.05 and +0.02, and 
B.C.=$-$0.15 and $-$0.24, respectively.  Comparing to the observed 
colors (Wilson 1992 photometry), colors have been dereddened using
R$_v$=3.1; also the intrinsic luminosity (L/L$_\odot$) has
been determined after adopting the distance modulus 
(23.49 $\pm$0.08) from Gallart \etal (1996a).    Luminosity and
reddening are listed in Table~\ref{basic}, along with the radii
determined from L/L$_\odot$ and \teff.

\section{Abundances \label{abundances}}

Elemental abundances were calculated using both spectrum
synthesis and individual line width analyses.   All calculations
were done using a modified and updated version of 
LINFOR\footnote{LINFOR was originally developed by H.\,Holweger,
W.\,Steffan, and W.\,Steenbock at Kiel University.   It has
been upgraded and maintained by M.\,Lemke, with additional
modifications by N.\,Przybilla.}.

In Table~\ref{lines1}, the atomic data adopted from the literature
are listed.   An attempt was made to adopt laboratory measurements 
(e.g., O'Brien \etal 1991 for \ion{Fe}{1})
and opacity project data 
(e.g., Bi\'emont \etal (1991) for \ion{O}{1}).
Critically examined data were selected next 
(e.g., NIST data from Fuhr, Martin, \& Wiese 1988 for \ion{Fe}{2}), 
followed by the semi-empirical values calculated by Kurucz (1988).
Solar abundances are from Grevesse \& Sauval (1998).

\begin{figure}
\plotone{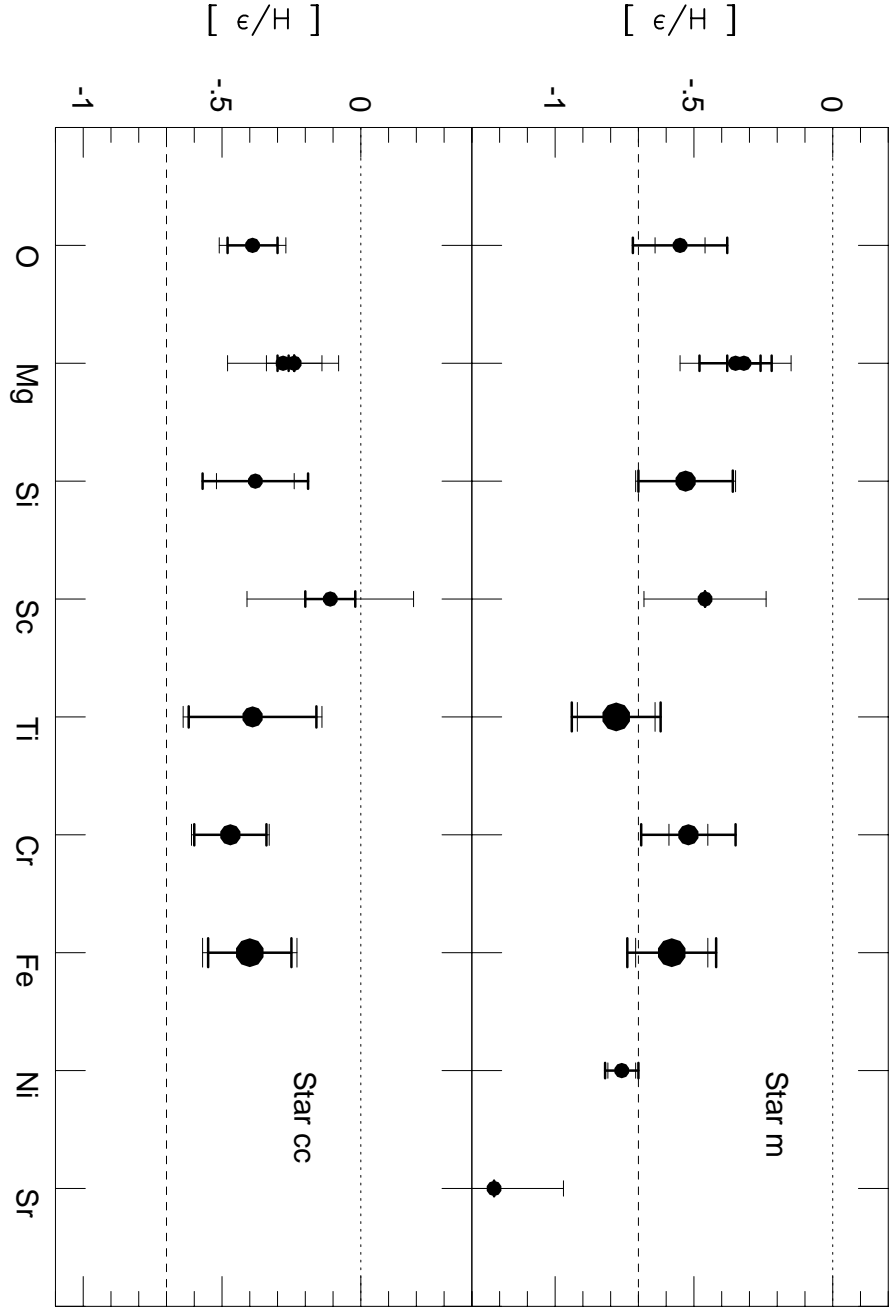}
\caption{Elemental abundances for Star m (top) and Star cc (bottom)
relative to the the Sun.   Two errorbars are shown for each point;
{\it thick line} for the line-to-line scatter, and {\it thin line}
for the systematic uncertainties (see Table~\ref{abu}).
The largest data points include $\ge$16 line abundances,
and the smallest include $\le$5.
The mean SMC underabundance is noted by the {\it dashed line}
(although note that the SMC A-supergiant Sr result is off the 
dashed line and this graph at -1.7).
\label{abund}}
\end{figure}

Averaged elemental abundances per star are listed in Table~\ref{abu}
and plotted in Figure~\ref{abund}.   Two error estimates are noted;
the first is the line-to-line scatter ($\sigma$), and the second is
a {\it estimated} systematic uncertainty based on uncertainties in
the atmospheric parameters.   Abundance uncertainties due to model 
atmosphere parameters are shown in Table~\ref{abu-unc}. 
The systematic uncertainty is probably an overestimate since we have 
simply added the possible uncertainties (in Table~\ref{abu-unc}) 
in quadrature, not accounting for the fact that some \teff-gravity 
combinations are excluded by the data 
(only for \ion{Mg}{1} did we account for the excluded combinations).
We also calculated the effect of the metallicity in the
model atmosphere calculation by scaling the opacity distribution
function by $-$0.4\,dex overall (or Z$_\odot$/3), but found this
affected the elemental abundances by $\le$0.03\,dex.  

\begin{figure}
\plotone{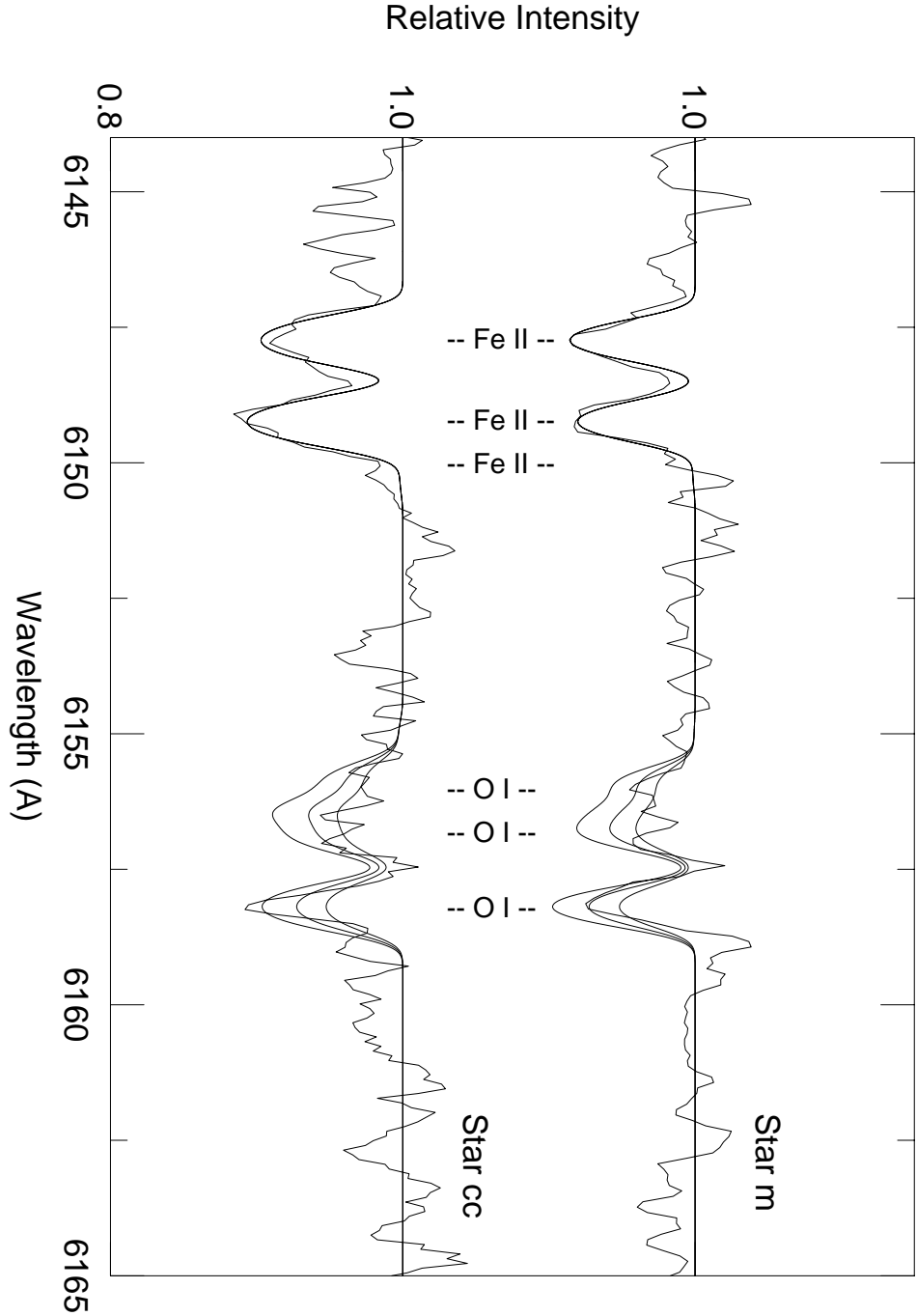}
\caption{Spectrum synthesis of the \ion{O}{1}~$\lambda$6155-6158 lines,
as well as the \ion{Fe}{2} lines near $\lambda$6148, for Star m (top)
and Star cc (bottom).   
Three oxygen abundances are shown, 12+log(O/H)=8.5, 8.7, 8.9.
The spectrum synthesis parameters are those listed in 
Table~\ref{basic}, and the iron line abundances are those 
listed in Table~\ref{lines1}.
\label{ospec}}
\end{figure}

Oxygen abundances are primarily from the spectrum synthesis of the 
\ion{O}{1} feature near 6158~\AA;  the spectra and LTE fits for both 
stars are shown in Figure~\ref{ospec}.   
Synthesis of three \ion{Fe}{2} lines ($\lambda\lambda$6147, 6149, 
and 6150\,\AA) were done simultaneously with the \ion{O}{1} synthesis.
Results for rotational velocities, radial velocities, and
macroturbulence were determined at this time and compared to
spectrum syntheses of the magnesium and iron lines near 
5200\,\AA\ and 4390\,\AA.   The synthesis parameters are listed 
in Table~\ref{basic}. 
Abundances per line from the spectrum synthesis are listed 
in Table~\ref{lines1}, and averaged for the best LTE oxygen abundance.
Non-LTE corrections for oxygen are taken 
from the detailed calculations by Przybilla \etal (2000); based on the 
model atmosphere parameters, the corrections are $-$0.2 and $-$0.3~dex 
for Star cc and Star m, respectively.  The final non-LTE oxygen 
abundances are listed in Table~\ref{abu}.

{\it Iron-group:}  The \ion{Fe}{2} results dominate our understanding
of the metallicities in these two stars.   This is because there are 
many spectral lines analysed which are relatively insensitive to
uncertainties in the atmospheric parameters 
(other than $\xi$, but that is {\it determined} from
\ion{Fe}{2} itself), and not expected to 
suffer significant non-LTE effects. 
\ion{Cr}{2} underabundances are in remarkably good 
agreement with those of \ion{Fe}{2} in both stars. 
The \ion{Ni}{2} abundance in Star m is slightly more underabundant
than \ion{Fe}{2}, but only two lines are sampled and the result
is within the 1\,$\sigma$ errors.
The difference of 0.2\,dex in the \ion{Fe}{2} abundance
between Star m and Star cc is interesting (discussed further
in Section~\ref{chem-today}),
but could simply reflect the uncertainties in the analyses. 

{\it Oxygen and the $\alpha$-elements:} The oxygen abundances in
both stars were determined from a combination of spectrum
synthesis and line analysis.   
For Star m, the spectrum synthesis favors a value near 
12+log(O/H)=8.5 $\pm$0.1 (LTE) for the blended 
6155-6156~\AA\ lines, yet synthesis and line analysis of the 
6158~\AA\ line  favors an abundance of 8.7 $\pm$0.1 (in LTE).   
We attribute this difference to the local S/N, since the lines 
are from the same multiplet 
(and have good atomic data which has been used to successfully 
determine the same line abundances of oxygen in Galactic stars 
in the past).  An average LTE abundance is 
12+log(O/H)=8.58 $\pm$0.17 ({\it $\pm$0.09}) 
(where the different abundances for the 6158~\AA\ line 
in Table~\ref{lines1} are themselves averaged a priori).
The spectrum synthesis of oxygen for Star cc is more problematic.
The synthesis parameters that best fit the two iron lines near 6148~\AA\
(and checked with magnesium line fits in the blue) do not produce
a spectrum that fits the \ion{O}{1}\,6158 line well.   Thus, there is 
a significant disparity between the line abundance from its equivalent 
width and that from spectrum synthesis.
We expect these difficulties arise from the S/N of the data at this
feature.  An average LTE abundance from the line abundances
in Table~\ref{lines1} is 12+log(O/H)=8.64 $\pm$0.09 ({\it $\pm$0.12}).
The final abundances listed in 
Table~\ref{abu} include the non-LTE corrections (discussed above). 

\ion{Si}{2} appears to be slightly more underabundant than 
the Mg abundances in these stars, although the same effect
was seen in Galactic A-type supergiants.
The \ion{Ti}{2} underabundance is similar to O and Fe in Star cc,
but it is $\sim$0.2\,dex lower than Fe in Star m (this is similar to
the \ion{Ni}{2} result, yet it is determined from many more lines
than Ni).   A simple adjustment of the atmospheric parameters
can bring the Ti result into better agreement with Fe,  
e.g., $\Delta\xi=-$1~\kms.
Another possibility is a neglected \ion{Ti}{2} non-LTE effect,
a preliminary non-LTE syntheses of \ion{Ti}{2} lines 
in A-supergiants (Becker 1998) suggests that the LTE abundances 
could be underestimated, as we have found.
\ion{Sc}{2} abundances are from very few lines, and we have
neglected the hyperfine structure terms in this analysis, thus
we shall not discuss those abundances further.

{\it s-process:} Only one line of \ion{Sr}{2} is clearly detected
and analysed in Star m ($\lambda$4077.7).
Its underabundance is much greater than that of \ion{Fe}{2}
([Sr/Fe]=$-$0.64 $\pm$0.00 ({\it $\pm$0.25}), which is
not surprising considering the large underabundances of Sr 
found in the Galactic and SMC A-type supergiants
([Sr/Fe]$\sim-$0.4 and $-$1.0, respectively). 
Non-LTE probably affects the absolute Sr abundances, thus 
differential abundances should be more accurate.   We find
the differential Sr abundances, i.e., Sr(Star m) $-$ Sr(SMC),
to be similar to the differential results for other elements.

\section{Discussion}

\subsection{Present-day Chemistry of NGC\,6822 \label{chem-today}}

For the first time, the present-day iron-group abundances have
been determined from stars in NGC\,6822.   The mean underabundance
is [Fe/H]=$-$0.49 $\pm$0.22 ({\it $\pm$0.21}).
This is from the \ion{Fe}{2} abundances, which are well supported 
by the results from the other iron-group elements, especially 
\ion{Cr}{2}, where the mean [Cr/H]=$-$0.50 $\pm$0.20 ({\it $\pm$0.16}).
In Table~\ref{diff}, all elemental abundances determined here are 
compared to solar, and differentially to the results from similar 
Galactic A-type supergiants (Venn 1995a,b) and SMC A-type supergiants 
(Venn 1999).  A comparison of the differential results, confirms 
that these two stars have a metallicity that is slightly higher 
than that in the SMC (also see Figure~\ref{abund}).

We also notice that the \ion{Fe}{2} abundance in Star cc is nearly
0.2~dex higher than in Star m.  A line-by-line differential abundance
comparison (from Table~\ref{lines1}) shows 
$\Delta$log(Fe/H)=+0.18 $\pm$0.08.  
It is difficult to gauge the significance of this 
difference at present; it is $>$2~$\sigma$, but also close 
to the estimated systematic errors for \ion{Fe}{2} from the
atmospheric analysis of each star ($\sim\pm$0.15).
 
The mean oxygen abundance determined from the two stars, 
12+log(O/H)=8.36 $\pm$0.19 ({\it $\pm$0.21}), is consistent with the nebular 
results, 8.25 $\pm$0.07 (Pagel \etal 1980, or 8.14 $\pm$0.08
considering only the three \hii regions where \ion{O}{3} was
detected).  Also, the two stellar oxygen abundances 
(8.28 and 8.44) are well within the range of the seven \hii region 
results (e.g., 12+log(O/H)=8.11 from association ``Hodge\,15'' and
8.50 from ``Hodge\,13''). 
In fact, we note that the two stars analysed here are located near the 
center of NGC\,6822, where Pagel \etal report oxygen of 8.44 $\pm$0.25.
That the oxygen abundance from the stars is consistent with
that from the nebulae is not surprising.   Similar results have
been found from recent analyses of B-stars in Orion 
(c.f., Cunha \& Lambert 1994), the Galactic oxygen gradient 
from B-stars (c.f., Rolleston \etal 2000, and references therein), 
and B-K supergiant
analyses in the SMC (c.f., Venn 1999, and references therein), 
M31 (Venn \etal 2000), 
and M33 (McCarthy \etal 1995, Monteverde \etal 2000).  

In Table~\ref{diff}, one can easily see that the mean 
oxygen abundance from these two stars in NGC\,6822 is less than solar, 
less than the Galactic A-type supergiants (this is significant since
the Sun appears to have a higher oxygen abundance than the nearby stars 
and nebulae, see Meyer \etal 1998), and more than in the SMC stars. 
Similarly, the oxygen and iron overabundances relative to the SMC
stars are nearly identical.
Notice also in this Table that the mean underabundance of oxygen
is the same as the mean underabundance of the other $\alpha$-elements  
(Mg, Si), when examined relative to the Galactic A-type supergiants.
Other elements (Ne \& Ar) examined by the nebular analyses 
show underabundances ranging from $-$0.2 to $-$0.6 
(Pagel \etal 1980, Skillman \etal 1989b), thus less than solar
but greater than (or similar) to SMC abundances,
consistent with the stellar results. 
Finally, Sr in Star m is less than solar, but considering the
differential abundances, then Sr appears to be significantly
overabundant relative to the Galactic and SMC A-supergiants.

\subsection{Chemical Evolution of NGC6822 \label{chem-evol}}

It is well known that the O/Fe ratio is a key constraint for
the chemical evolution model of a galaxy.   This is because
O is primarily synthesized in Type II supernovae, whereas most
of the Fe comes from Type Ia supernovae; thus a burst of star 
formation temporarily increases the O/Fe ratio in the ISM 
as massive stars quickly enrich the local ISM in O,
while lower mass stars slowly increase the amount of Fe 
(Wheeler, Sneden, \& Truran 1989, Gilmore \& Wyse 1991).   

We find a mean ratio of [O/Fe]=+0.02 $\pm$0.20 ({\it $\pm$0.21}) 
for the two stars analysed here. 
As seen in Table~\ref{diff}, our mean [O/Fe] ratio is similar 
to that of the SMC A-type supergiants (also note that the 
SMC A-supergiant results are in very good agreement with abundances 
from F-K supergiants, B-stars, and \hii regions, see the discussion
on differential SMC abundances by Venn 1999).  This O/Fe ratio
is also similar to that found from stars in the LMC
(see Hill 1997, 1999, Hill \etal 1995, Russell \& Bessell 1989, 
Luck \etal 1998, and Luck \& Lambert 1992).
This O/Fe ratio may also be similar
to that of the Galactic disk stars at [Fe/H]=$-$0.5; 
Edvardsson \etal (1993) show that [O/Fe] ranges from 
$\sim$+0.1 to +0.25 in Galactic F-G disk dwarfs at 
[Fe/H]=$-$0.5, which is slightly higher than our ratio but within
our error range. 
  
Our other mean $\alpha$-elements-to-iron ratio is similar to 
both the SMC A-supergiant results and the Galactic F-G disk
dwarfs.   In Table~\ref{diff}, the mean 
$\alpha$(Mg,Si)/Fe = +0.13 $\pm$0.20 ({\it $\pm$0.26}), while
the mean from the SMC A-supergiants is +0.03 $\pm$0.20. 
For the Galactic F-G disk dwarfs, Edvardsson \etal (1993) 
show that [Mg,Ti/Fe] ranges from +0.1 to +0.3 at [Fe/H]=$-$0.5 
(though [Si/Fe] looks slightly flatter, ranging from +0.05 to +0.2).  
This range overlaps the SMC results, and our new NGC\,6822 results.

Analytical chemical evolution models for the Magellanic Clouds 
have been published by Pagel \& Tautvai\u{s}ien\.e (1998).  These models 
show the predicted abundance ratios over a range of iron abundances.  
Our [O/Fe] and [Fe/H] results suggest that their models may also 
be appropriate for NGC\,6822, even though the star formation histories 
in the Clouds are somewhat different from that determined in NGC\,6822
by Gallart \etal (1996b,c).   For example, a comparison of Figure~12
in Gallart \etal (1996c) and the bursting model of Figure~2 in 
Pagel \& Tautvai\u{s}ien\.e (1998) suggests that both galaxies have
undergone recent bursts of star formation after a hiatus at 
intermediate ages\footnote{The star formation history of the SMC is
currently controversial, particularly with respect to a hiatus in
the star formation rate at intermediate epochs.   For example, 
the ages and metallicities of intermediate-aged clusters can be 
interpreted as supporting either a smooth star formation rate 
(e.g., Da\,Costa \& Hatzidimitriou 1998), or a bursting star formation
history with an intermediate-aged hiatus (Mighell \etal 1998). 
Most dynamical models of the SMC, as well as the interpretation of 
the O/Fe ratio in the SMC, do suggest some form of a bursting 
star formation history, most likely related to interactions with
the LMC and Galaxy (c.f., review by van den Bergh 2000, Chapter~7).}.   
However, the strength and timing of the recent 
bursts, and the details of star formation in the older populations, 
are quite different.

Finally, our [Sr/Fe] ratio in Star m is interesting since 
the s-process elements are thought to form primarily by thermal-pulsing 
in intermediate-mass AGB stars.   Thus, Sr samples a unique mass 
range (from O and Fe), an asset when reconstructing the star formation 
history of the galaxy. 
This can also be seen by the differences in the predicted Sr/Fe ratios 
in the Pagel \& Tautvai\u{s}ien\.e (1998) models for the Magellanic Clouds. 
Our one Sr/Fe ratio, [Sr/Fe]=$-$0.64 $\pm$0.00 ({\it $\pm$0.25}), 
suggests that Sr is slightly less underabundant than in the A-F
supergiants in the SMC, consistent with the results from other elements. 

It is possible to observe additional elements in A-F supergiants,
e.g., C, N, \& S with observations at wavelengths from 7000 to 9500\,\AA.  
Nitrogen abundances would be interesting considering the very low nebular 
N abundances in NGC\,6822 (Pagel \etal 1980, Kobulnicky \& Skillman 1998), 
however surface N in A-supergiants is very likely to be enhanced through 
mixing with interior gas.

\subsection{Spatial Abundance Variations \label{spatial}} 

The two stars analysed here are located near the center of
NGC\,6822, and have similar elemental abundances to each other.     
The two stars do not span a significant range
in locations in NGC\,6822 to examine spatial abundance variations
on their own, therefore we have calculated the galactocentric
distances of these stars to compare to nebular abundances.

To calculate the NGC\,6822 galactocentric distances, we have adopted
the \ion{H}{1} dynamical center coordinates (19$^h$ 42$^m$ 06.7$^s$
and $-$14$^o$ 55' 22.0'', 1950), position angle (112$^o$), 
and inclination angle (inner part, 50.1$^o$) from 
Brandenburg \& Skillman (1998).   Our distances and the oxygen 
abundances for our two stars, Pagel {\it et~al.}'s \hii regions, and
two planetary nebulae from Richer \& McCall (1995) are 
listed in Table~\ref{dist}.   We include the results from the
two bright planetary nebulae considering that Richer (1993) has
shown that the bright planetary nebulae in the Magellanic Clouds
yield identical mean oxygen abundances as the \hii regions
(also see the discussion in Richer \& McCall 1995).  

\begin{figure}
\plotone{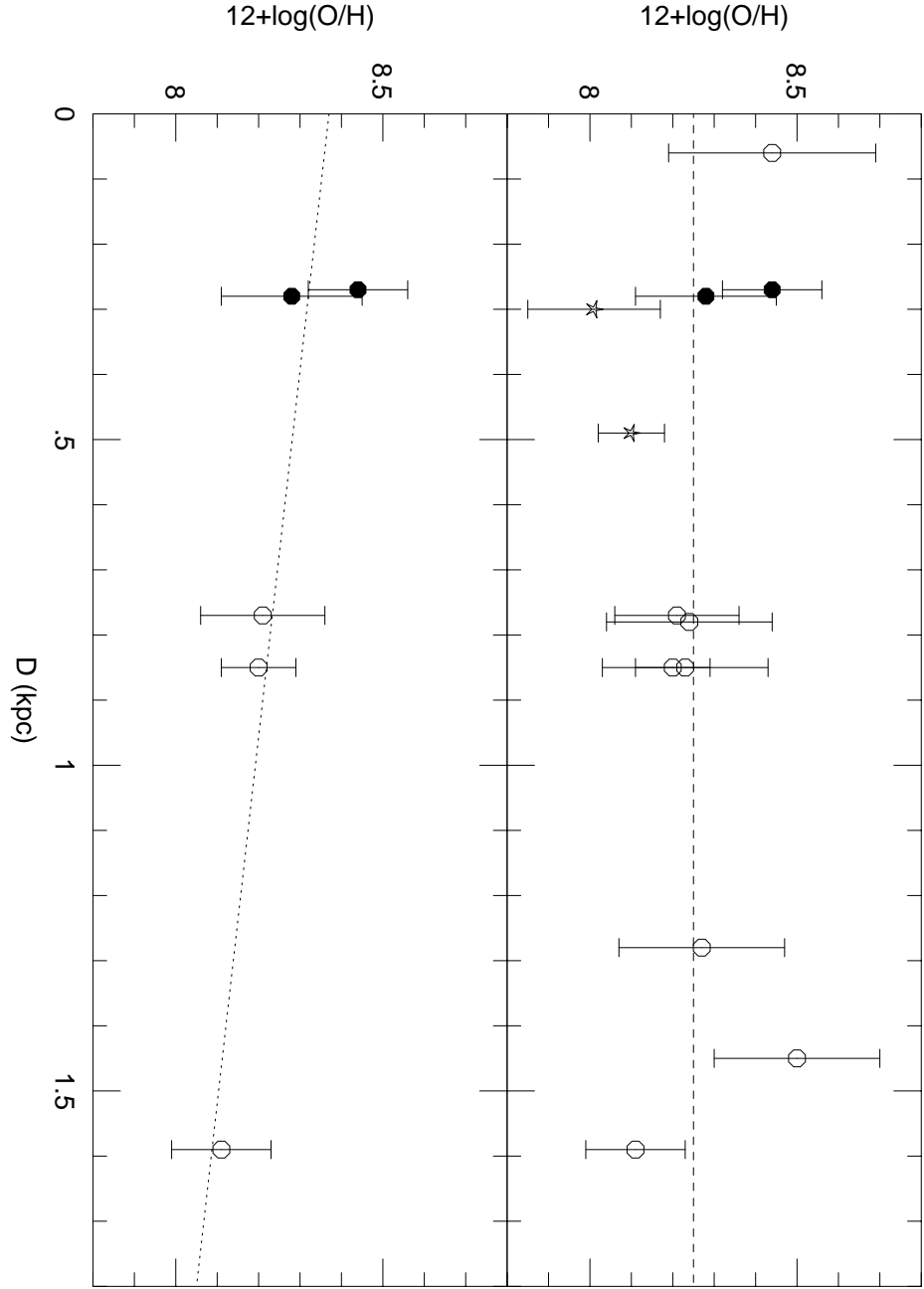}
\caption{Oxygen abundances versus NGC\,6822 galactocentric distance.
{\it Filled circles} represent the stellar data, {\it hollow circles} 
show the Pagel \etal (1980) nebular data, and {\it hollow stars} note 
two planetary nebulae results from Richer \& McCall (1995).   
The {\it dashed line} in the top panel shows the mean oxygen abundance 
from Pagel \etal (12+log(O/H)=8.25).
In the bottom panel, only the stellar 
abundances and those from \hii regions where \ion{O}{3} was detected 
are shown.   A least squares fit to the data ({\it dotted line})
suggests a slope of $-$0.18 dex/kpc.
\label{sgrad}}
\end{figure}

When all of the data are examined together, we find no evidence
for an abundance gradient in NGC\,6822 (see Figure~\ref{sgrad}a).
Pagel \etal (1980) came to the same conclusion from examination 
of the \hii region data alone, although they were not aware of 
the orientation of the disk of NGC\,6822. 
However, when only the most reliable of the data for the present-day
oxygen abundance are examined, i.e., the stellar data and the three \hii 
regions where \ion{O}{3}~$\lambda$4363 was detected 
(thus better electron temperatures), then a trend does emerge.   
There is the signature of an oxygen gradient with a slope    
near $-$0.18 dex/kpc (see Figure~\ref{sgrad}b).

At this time, we do not suggest that we have found an actual abundance
gradient in NGC\,6822;  we simply report that a selective subset of the 
data is consistent with a trend.   As seen in Figure~\ref{sgrad}a,
the data are also consistent with no trend at all.   New spectroscopy
of stars at larger distances and/or the \hii regions could address this
question.  (It may be worth noting that these three \hii regions do
span across the full extent of the northern half of NGC\,6822). 
A significant abundance gradient in NGC\,6822 would be surprising,
particularly one that is $\sim$3x larger than that seen in the Galaxy
(c.f., Rolleston \etal 2000, Shaver \etal 1983).
Several other dwarf irregular galaxies have abundances from multiple
\hii regions, and yet, with the exception of NGC\,5253, 
there are no cases of significant internal
chemical fluctuations or abundance gradients 
(see Kobulnicky \& Skillman 1997 and references therein).  
The lack of abundance variations in dwarf irregular galaxies has 
been interpreted as evidence against in situ enrichment
(e.g., the instantaneous recycling approximation used in 
galaxy chemical evolution models), and thus a counter example 
would be intriguing.

\subsection{Reddening and Cepheid Distances}

Foreground reddening to NGC\,6822 is significant because of its low
galactic latitute ($\sim-$18.4$^o$).  
From our model atmospheres, the intrinsic (B-V)$_o$ colors for our 
two stars were determined, and the reddening found to be
E(B-V)=+0.29 and +0.38 for Star~cc and Star~m,
respectively (see Table~\ref{basic}).
Variable reddening has been found by several authors across
NGC\,6822.   For example, Massey \etal (1995) found a systematic
spatial trend from E(B-V)=0.26 near the eastern and western most edges,
and up to 0.45 near the bar of recently formed stars; other values 
of the mean (foreground +internal) reddening have ranged from 
$\sim$0.2 (Gallart \etal 1996a, Kayser 1967, Hodge 1977) to 
$\sim$0.45 (Wilson 1992, van den Bergh \& Humphreys 1979).
The values found for the two stars in this paper fall within this 
range, and are in good agreement with Massey \etal's results since 
Star~cc (near the eastern most edge) has a lower reddening than 
Star~m (centrally located along the star-forming bar). 

We have examined the positions of the eight Cepheid variables 
studied by Gallart \etal (1996a), who found a mean reddening 
of E(B-V)=+0.24 $\pm$0.03 to derive a true distance modulus 
to NGC\,6822 of ($m-M$)$_o$=23.49 $\pm$0.08.  
Four Cepheid variables (V1, V2, V3, and V13) are within 1' 
of our supergiant targets, but only one Cepheid is truly near
one of our targets; V3 is located 9'' ($\sim$50 pc) from Star cc.  
Adopting the Star~cc reddening value, we predict the reddening 
for V3 is $\sim$0.05 magnitudes higher than the Gallart \etal 
mean value, yielding a slightly smaller distance modulus 
(although this difference is certainly 
within Gallart \etal's errors).    
More interesting is the Cepheid V2, which appears to be 
centrally located in the star-forming bar, and is only 23'' 
($\sim$0.2 kpc) from Star~cc and 34'' ($\sim$0.3 kpc) 
from Star~m.   If this star has the higher reddening
attributed to stars in the central star-forming region,
and supported by our Star~m reddening result,
then its reddening could be underestimated by as much
as 0.2 magnitudes and its distance overestimated by 
$\sim$10\% (45 kpc). 
 
As Gallart \etal (1996a) discuss, the large and 
variable reddening to NGC\,6822 is one of the largest
problems in determining its true distance.   
A combination of spectroscopic stellar reddening 
measurements and new photometric monitoring
of the known Cepheids could provide an improved 
distance to NGC\,6822.

\section{Conclusions and Future Work} 

We have presented the first stellar abundances in NGC\,6822,
which include O, other $\alpha$-elements, and iron-group
abundances (plus one s-process element).  Comparison of the
oxygen abundances to those from \hii regions shows that the
stellar abundances are in excellent agreement with the nebulae.
This has allowed us to examine the present-day O/Fe ratio in NGC\,6822 
for the first time.   We find the mean 
[O/Fe]=+0.02 $\pm$0.20 ({\it $\pm$0.21}) is more similar to that in
the SMC, than in metal-poor Galactic stars.
We have also used the oxygen data to search for spatial abundance
variations.  Calculating NGC\,6822 galactocentric distances using
disk parameters from \ion{H}{1} 21\,cm synthesis observations 
shows no oxygen gradient when
all data are considered.   However, when only a subset of the most
reliable oxygen abundances is used, the data are
also consistent with a gradient of slope $-$0.18 dex/kpc.  

We recommend additional stellar and nebular spectroscopic analyses
of NGC\,6822 in the future to address the following questions;

(1) Is there an oxygen abundance gradient in this dwarf irregular galaxy?

(2) Is there a range in the iron-group abundances? 

Finally, similar analyses of stars in other Local Group dwarf irregular
galaxies will give us new information on O/Fe ratios for a variety
of metallicities, and thus their star formation histories. 
Understanding the most metal-poor dwarf irregulars may be relevant 
to studies of high-z galaxies, which are likely to be similarly 
metal-poor, pre-merger galaxies.

\acknowledgments

KAV would like to acknowledge financial support from the National 
Science Foundation, CAREER grant AST-9984073, the Henry Luce Foundation 
through a Clare Boothe Luce Professorship award, and from Macalester College.
Also, KAV would like to thank Lissa Miller and Denis Foo Kune for research 
assistance, Carme Gallart for providing the Cepheid positions, and the
referee B.\,E.\,Reddy for several helpful comments that improved this
manuscript.
JKM would like to thank the staff of the W.\,M.\,Keck Observatory,
and in particular observing assistants Terry Stickel and Wayne Wack
for their efforts on the summit in support of the Keck HIRES
observations.
SJS acknowledges financial support from the PPARC.
EDS acknowledges financial support from a NASA LTSA grant,
NAG5-9221.


\clearpage
\begin{deluxetable}{lrr}
\footnotesize
\tablecaption{Coordinates and Color Information for Stars in NGC\,6822
	\label{basic}} 
\tablewidth{0pt}
\tablehead{
\colhead{} & \colhead{Star cc} & \colhead{Star m}   
} 
\startdata
Names            & A13   & A101   \nl
		 & CW185 & CW173  \nl
\nl
$\alpha$ (J2000)\tablenotemark{a} &  19 44 53.4 &  19 44 56.5 \nl
$\delta$ (J2000)\tablenotemark{a} & $-$14 46 42   & $-$14 46 14 \nl
Hodge OB\#\tablenotemark{a}    & 10          & \nodata         \nl
V\tablenotemark{a}      & 17.36 $\pm$0.01 & 17.38 $\pm$0.03 \nl
(B-V)\tablenotemark{a}  & +0.34           & +0.40 \nl
{\it OLD Sp.~Ty.}\tablenotemark{b} & {\it B8-A0 I} & {\it A0 I} \nl
{\it OLD  V}\tablenotemark{b}      & {\it 16.97} & {\it 17.04} \nl
{\it OLD (B-V)}\tablenotemark{b}   & {\it +0.36} & {\it +0.20} \nl 
{\it OLD E(B-V)}\tablenotemark{b}  & {\it +0.36} & {\it +0.20} \nl 
\nl
NEW Sp.~Ty. &  A3 Ia   &  A2 Ia \nl
\teff (K)   &  8500 $\pm$150 & 9000 $\pm$150 \nl
\logg       &  1.1 $\pm$0.1 & 1.3 $\pm$0.1  \nl
E(B-V) & +0.29 & +0.38 \nl 
log(L/L$_\odot$) & 4.78 & 4.92 \nl
R/R$_\odot$      & 113  & 119  \nl 
$\xi$ (\kms) &   6 $\pm$1 &   6 $\pm$1 \nl
$v$sin$i$ (\kms)      &  28 $\pm$2 &  15 $\pm$1 \nl
RV (\kms)             & $-$55 $\pm$2 & $-$65 $\pm$2 \nl
\enddata
\tablecomments{Names A\# from Kayser 1967, CW\# from Wilson 1992.}
\tablenotetext{a}{Magnitudes and colors from Wilson 1992,
  (also, OB association \#, originally identified by Hodge 1977);
  coordinates from erratum (Wilson 1995).}
\tablenotetext{b}{Magnitudes and colors from 
  van den Bergh \& Humphreys 1979.  Their reddening estimates 
  adopt (B-V)$_o$=0.0 for both stars.}
\end{deluxetable}


\clearpage
 
\begin{deluxetable}{rrrrrllrlrl}
\footnotesize
\tablecaption{Line Strengths, Atomic Data and Abundances \label{lines1}}
\tablewidth{0pt}
\tablehead{
\colhead{Elem} & \colhead{RMT} & \colhead{$\lambda$ (\AA)} &  
\colhead{$\chi$ (eV)} & \colhead{log~gf} & 
\colhead{REF\tablenotemark{\dagger}} & \colhead{accy} &
\colhead{Star cc} & \colhead{Log$\epsilon$(cc)$_{\rm LTE}$} & 
\colhead{Star m}  & \colhead{Log $\epsilon$(m)$_{\rm LTE}$}
} 
\startdata
 800 & 10 & 6155.99 & 10.74 & -0.67 & op  &    &   S     & 8.6     &  S      & 8.5      \nl
 800 & 10 & 6156.78 & 10.74 & -0.45 & op  &    &   S     & 8.6     &  S      & 8.5      \nl
 800 & 10 & 6158.19 & 10.74 & -0.31 & op  &     & 61/S   & 8.62/8.8 & 66/S   & 8.77/8.7 \nl
1200 &  3 & 3829.36 &  2.71 & -0.21 & fw  & B  & \nodata & \nodata & 31      & 6.85    \nl
1200 &  2 & 5172.68 &  2.71 & -0.38 & fw  & B  &  95     & 7.09    & 40      & 7.07    \nl
1200 &  2 & 5183.60 &  2.72 & -0.16 & fw  & B  & 125     & 7.08    & 61      & 7.08    \nl
1201 &  5 & 3848.21 &  8.86 & -1.56 & fw  & C  & \nodata & \nodata & 33      & 7.26    \nl 
1201 & 10 & 4390.57 & 10.00 & -0.50 & fw  & D  &  71     & 7.31    & 60      & 7.19    \nl
1401 &  1 & 3853.66 &  6.86 & -1.60 & fw  & E  & \nodata & \nodata & 133     & 7.20   \nl
1401 &  1 & 3862.59 &  6.86 & -0.90 & fw  & D+ & \nodata & \nodata & 210     & 7.26   \nl
1401 &  3 & 4128.07 &  9.84 &  0.31 & fw  & C  & \nodata & \nodata & 114     & 6.81   \nl
1401 &  3 & 4130.89 &  9.84 &  0.46 & fw  & C  & \nodata & \nodata & 133     & 6.86   \nl
1401 &  5 & 5041.02 & 10.07 &  0.17 & fw  & D+ & 118     & 7.37    & 103     & 7.09   \nl
1401 &  4 & 5957.56 & 10.07 & -0.35 & fw  & D  & 37      & 6.99    & 45      & 7.02   \nl
1401 &  4 & 5978.93 & 10.07 & -0.06 & fw  & D  & \nodata & \nodata & 63      & 7.00   \nl 
2101 & 15 & 4314.08 &  0.62 & -0.05 & k88 &    & 137     & 2.92    & 38      & 2.64   \nl
2101 & 15 & 4320.73 &  0.61 & -0.21 & k88 &    & 135    & 3.06    & $\le$20 & \nodata \nl
2201 & 72 & 3741.63 &  1.58 & -0.11 & mfw & D  & \nodata & \nodata & 72      & 3.77   \nl
2201 & 34 & 3900.56 &  1.13 & -0.44 & mfw & D  & \nodata & \nodata & 90      & 3.93   \nl
2201 & 34 & 3913.48 &  1.12 & -0.53 & mfw & D  & \nodata & \nodata & 98      & 4.08   \nl
2201 & 11 & 4012.40 &  0.57 & -1.61 & mfw & C  & \nodata & \nodata & 47      & 4.35   \nl
2201 & 105& 4163.63 &  2.59 & -0.40 & mfw & D  & \nodata & \nodata & 48      & 4.44   \nl
2201 & 105& 4171.92 &  2.60 & -0.56 & mfw & D  & \nodata & \nodata & 30      & 4.37   \nl
2201 & 41 & 4290.22 &  1.16 & -1.12 & mfw & D- & \nodata & \nodata & 55      & 4.30   \nl
2201 & 20 & 4294.09 &  1.08 & -1.11 & mfw & D- & \nodata & \nodata & 36      & 4.02   \nl
2201 & 41 & 4300.06 &  1.18 & -0.77 & mfw & D- & \nodata & \nodata & 91      & 4.27   \nl
2201 & 41 & 4301.92 &  1.16 & -1.16 & mfw & D- & \nodata & \nodata & 35      & 4.11   \nl
2201 & 41 & 4307.90 &  1.16 & -1.29 & mfw & D- & {\it 220} & \nodata & 41      & 4.32   \nl
2201 & 41 & 4312.86 &  1.18 & -1.16 & mfw & D- & 151     & 4.67    & 40      & 4.19   \nl
2201 & 41 & 4314.97 &  1.16 & -1.13 & mfw & D- & 88      & 4.20    & 36      & 4.09   \nl
2201 & 19 & 4395.03 &  1.08 & -0.66 & fmw & D- & 200     & 4.44    & 87      & 4.05   \nl
2201 & 51 & 4399.77 &  1.24 & -1.27 & fmw & D- & 116     & 4.58    & 21      & 4.02   \nl
2201 & 19 & 4443.80 &  1.08 & -0.70 & fmw & D- & 196     & 4.44    & 83      & 4.06   \nl
2201 & 19 & 4450.48 &  1.08 & -1.45 & mfw & D- & 89      & 4.46    & $\le$20 & \nodata \nl
2201 & 31 & 4501.27 &  1.12 & -0.75 & mfw & D- & 142     & 4.13    & 77      & 4.09   \nl
2201 & 50 & 4563.76 &  1.22 & -0.96 & mfw & D- & 151     & 4.47    & \nodata & \nodata \nl
2201 & 82 & 4571.96 &  1.57 & -0.52 & mfw & D- & 191     & 4.55    & \nodata & \nodata \nl
2201 & 92 & 4805.09 &  2.06 & -1.12 & fmw & D- & 81      & 4.72    & 20      & 4.36   \nl
2201 &114 & 4874.01 &  3.09 & -0.79 & mfw & D  & 46      & 4.77    & $\le$20 & \nodata \nl
2201 & 70 & 5154.07 &  1.57 & -1.92 & mfw & D- & 52      & 4.91    & $\le$20 & \nodata \nl
2201 & 70 & 5226.54 &  1.57 & -1.29 & mfw & D- & 136     & 4.90    & b & \nodata \nl
2401 &18-26& 4111.00 & 3.74 & -1.92 & k88 &    & \nodata & \nodata & 39      & 5.43   \nl
2401 & 31 & 4242.36 &  3.87 & -1.17 & sl  &    & \nodata & \nodata & 53      & 4.92   \nl
2401 & 31 & 4261.91 &  3.87 & -1.34 & sl  &    & \nodata & \nodata & 56      & 5.12   \nl
2401 & 31 & 4275.57 &  3.86 & -1.52 & sl  &    & \nodata & \nodata & 47      & 5.19   \nl
2401 & 31 & 4284.19 &  3.86 & -1.67 & sl  &    & \nodata & \nodata & 34      & 5.18   \nl
2401 & 44 & 4555.00 &  4.07 & -1.30 & sl  &    & 76      & 5.20    & \nodata & \nodata \nl
2401 & 44 & 4616.63 &  4.07 & -1.36 & sl  &    & 63      & 5.15    & \nodata & \nodata \nl
2401 & 44 & 4618.80 &  4.07 & -0.84 & sl  &    & 139     & 5.21    & \nodata & \nodata \nl
2401 & 44 & 4634.10 &  4.07 & -0.99 & sl  &    & 105     & 5.11    & \nodata & \nodata \nl
2401 & 30 & 4812.34 &  3.86 & -1.96 & sl  &    & 29      & 5.10    & $\le$25 & \nodata \nl
2401 & 30 & 4824.13 &  3.87 & -0.97 & sl  &    & 181     & 5.50    & 107     & 5.15   \nl
2401 & 30 & 4848.24 &  3.86 & -1.15 & sl  &    & 118     & 5.22    & 90      & 5.19   \nl
2401 & 30 & 4876.41 &  3.86 & -1.46 & sl  &    & 98      & 5.38    & 60      & 5.25   \nl
2401 & 43 & 5237.33 &  4.06 & -1.16 & mfw & D  & 89      & 5.14    & 66      & 5.13   \nl
2401 & 43 & 5274.96 &  4.05 & -1.29 & k88 &    & 64      & 5.06    & 56      & 5.16   \nl
2600 & 21 & 3787.88 &  1.01 & -0.84 & ob  &    & \nodata & \nodata & 21      & 7.35   \nl
2600 & 20 & 3820.43 &  0.86 &  0.16 & ob  &    & \nodata & \nodata & 80      & 6.96   \nl
2600 & 20 & 3825.88 &  0.91 & -0.03 & ob  &    & \nodata & \nodata & 47      & 6.87   \nl
2600 &  4 & 3859.91 &  0.00 & -0.71 & fmw & B+ & \nodata & \nodata & 50      & 7.00   \nl
2600 & 43 & 4063.60 &  1.56 &  0.06 & ob  &    & \nodata & \nodata & 29      & 6.93   \nl
2600 & 42 & 4325.76 &  1.61 &  0.01 & ob  &    & 83      & 7.11    & b & \nodata \nl
2600 & 41 & 4383.54 &  1.48 &  0.21 & ob  &    & 104     & 6.97    & b & \nodata \nl
2600 & 41 & 4404.75 &  1.56 & -0.15 & ob  &    & 59      & 7.03    & $\le$20 & \nodata \nl
2600 &318 & 4891.50 &  2.85 & -0.11 & ob  &    & 31      & 7.50    & $\le$20 & \nodata  \nl
2601 & 14 & 3783.35 &  2.27 & -3.16 & k88 &    & \nodata & \nodata & 110     & 6.95   \nl
2601 &  3 & 3945.21 &  1.70 & -4.25 & fmw & D  & \nodata & \nodata & 53      & 7.19   \nl
2601 & 28 & 4122.64 &  2.58 & -3.38 & fmw & D  & \nodata & \nodata & 50      & 6.82   \nl
2601 & 27 & 4173.45 &  2.58 & -2.18 & fmw & C  & \nodata & \nodata & 162     & 6.55   \nl
2601 & 28 & 4178.86 &  2.58 & -2.48 & fmw & C  & \nodata & \nodata & 170     & 6.91   \nl
2601 & 28 & 4258.15 &  2.70 & -3.40 & fmw & D  & \nodata & \nodata & 54      & 6.95   \nl
2601 & 27 & 4273.32 &  2.70 & -3.34 & fmw & D  & \nodata & \nodata & 48      & 6.82   \nl
2601 & 28 & 4296.57 &  2.70 & -3.10 & mfw & D  & \nodata & \nodata & 82      & 6.82   \nl
2601 & 27 & 4303.17 &  2.70 & -2.49 & fmw & D  & \nodata & \nodata & 176     & 7.04   \nl
2601 & 27 & 4351.77 &  2.70 & -2.10 & fmw & C  & {\it 280} & \nodata & 208     & 6.93   \nl
2601 & 27 & 4385.38 &  2.78 & -2.57 & fmw & D  & 188     & 7.08    & 123     & 6.74   \nl
2601 & 27 & 4416.83 &  2.78 & -2.61 & fmw & D  & 160     & 6.89    & 130     & 6.83   \nl 
2601 & 37 & 4472.92 &  2.84 & -3.43 & fmw & D  & 96      & 7.28    & 55      & 7.07   \nl
2601 & 37 & 4489.18 &  2.83 & -2.97 & fmw & D  & 138     & 7.12    & 99      & 6.99   \nl
2601 & 37 & 4491.40 &  2.86 & -2.69 & fmw & C  & 155     & 6.99    & 110     & 6.81   \nl
2601 & 38 & 4508.29 &  2.86 & -2.22 & fmw & D  & {\it 228} & \nodata & 193     & 7.00   \nl
2601 & 37 & 4515.34 &  2.84 & -2.48 & fmw & D  & {\it 230} & \nodata & 160     & 6.97   \nl
2601 & 37 & 4520.22 &  2.81 & -2.61 & fmw & D  & 175     & 7.02    & 133     & 6.87   \nl
2601 & 38 & 4541.52 &  2.86 & -3.05 & fmw & D  & 91      & 6.87    & \nodata & \nodata \nl
2601 & 38 & 4576.34 &  2.84 & -3.04 & fmw & D  & 118     & 7.05    & \nodata & \nodata \nl
2601 & 37 & 4582.84 &  2.84 & -3.09 & fmw & C  & 102     & 6.98    & \nodata & \nodata \nl
2601 &186 & 4635.32 &  5.96 & -1.65 & fmw & D- & 66      & 7.21    & \nodata & \nodata \nl
2601 & 43 & 4656.98 &  2.89 & -3.63 & fmw & E  & 80      & 7.38    & b  & \nodata \nl
2601 & 37 & 4666.75 &  2.83 & -3.33 & fmw & D  & 90      & 7.12    & 64      & 7.05   \nl
2601 & 43 & 4731.45 &  2.89 & -3.37 & fmw & D  & 94      & 7.23    & 53      & 7.02   \nl
2601 & 49 & 5197.57 &  3.23 & -2.10 & fmw & C  & 205     & 7.00    & 140     & 6.68   \nl
2601 & 49 & 5234.62 &  3.22 & -2.05 & fmw & C  & {\it 222} & \nodata & 167     & 6.83   \nl
2601 & 48 & 5264.81 &  3.23 & -3.19 & fmw & D  & \nodata & \nodata & 37      & 6.85   \nl
2601 &185 & 5272.40 &  5.96 & -2.03 & fmw & D  & \nodata & \nodata & 33      & 7.26   \nl
2601 & 44 & 5276.00 &  3.20 & -1.95 & fmw & C  & {\it 231} & \nodata & 170     & 6.74   \nl 
2601 & 41 & 5284.10 &  2.89 & -3.19 & fmw & D  & 106     & 7.13    & 56      & 6.87   \nl
2601 & 48 & 5362.86 &  3.20 & -2.74 & k88 &    & 148     & 7.19    & 91      & 6.92   \nl
2601 & 49 & 5425.25 &  3.20 & -3.36 & fmw & D  & \nodata & \nodata & 37      & 7.00   \nl
2601 &    & 5506.20 & 10.52 &  0.95 & fmw & D  & \nodata & \nodata & 32      & 6.88   \nl
2601 & 55 & 5534.85 &  3.24 & -2.92 & fmw & D  & \nodata & \nodata & 77      & 7.01   \nl
2601 & 74 & 6147.74 &  3.89 & -2.46 & fmwy* & D & 91     & 6.97    & 68      & 6.89   \nl
2601 & 74 & 6149.24 &  3.89 & -2.77 & fmw* & D & 100     & 7.35    & 60      & 7.12   \nl
2601 & 74 & 6238.38 &  3.89 & -2.48 & k88 &    & 89      & 6.97    & 40      & 6.61 \nl
2601 & 74 & 6247.56 &  3.89 & -2.36 & fmw* & D & 140     & 7.24    & 105     & 7.11   \nl         
2601 & 74 & 6416.91 &  3.89 & -2.70 & fmw* & D & \nodata & \nodata & 60      & 7.06   \nl
2801 & 11 & 3849.55 &  4.01 & -1.88 & k88 &    & \nodata & \nodata & 46      & 5.45   \nl
2801 & 11 & 4067.03 &  4.03 & -1.84 & k88 &    & \nodata & \nodata & 56      & 5.54   \nl
3801 &  1 & 4077.71 &  0.00 &  0.15 & k88 &    & \nodata & \nodata & 14      & 1.71   \nl
\enddata
\tablenotetext{\dagger}{Reference Key:
fw = Fuhr \& Wiese 1998,
fmw = Fuhr \etal 1988, 
fmwy = Fuhr \etal 1981, 
k88 = Kurucz 1988 (CD-18),
mfw = Martin \etal 1988, 
ob = O'Brien \etal 1991,
op = Hibbert \etal 1991, 
sl = Sigut \& Landstreet 1990,
wsm = Wiese, Smith, \& Miles 1969.
Some FeII adjusted for the lower solar FeII abundance than
used in the original reference (log$gf-$0.15 denoted by *). 
Capital letters denote estimated accuracy 
(where uncertainties of $gf$ values are within 
3\%=A, 10\%=B, 25\%=C, 50\%=D).  }
\end{deluxetable}


\clearpage
\begin{deluxetable}{lclllcc}
\footnotesize
\tablecaption{Magnesium Line NLTE corrections \label{mgnlte}}
\tablewidth{0pt}
\tablehead{
\colhead{$\lambda$} & \colhead{Levels} & 
\colhead{$\chi$} & \colhead{log($gf$)} & \colhead{REF} &
\colhead{Star cc} & \colhead{Star m} \\[.2ex] 
\colhead{(\AA)} & \colhead{} &
\colhead{(eV)} & \colhead{} & \colhead{} &
\colhead{EQW \ \ \ LTE \ \ \ NLTE} & \colhead{EQW \ \ \ LTE \ \ \ NLTE} 
} 
\startdata
\ion{Mg}{1} \nl
3829.36 & 3$p^3$P$^0$ $-$  3$d^3$D  &  2.71 & -0.21 & fwB & \nodata & 
	31 \ \ \ \ \ 6.85 \ \ \ \ \ 7.08 \nl
5172.68 & 3$p^3$P$^0$ $-$  4$s^3$S  &  2.71 & -0.38 & fwB & 
	\ \ 95 \ \ \ \ \  7.11 \ \ \ \ \ 7.31 & 
	40 \ \ \ \ \ 7.07 \ \ \ \ \ 7.30 \nl 
5183.60 & 3$p^3$P$^0$ $-$  4$s^3$S  &  2.71 & -0.16 & fwB & 
	125 \ \ \ \ \ 7.10 \ \ \ \ \ 7.29 & 
	60 \ \ \ \ \ 7.07 \ \ \ \ \ 7.30 \nl
\nl
\ion{Mg}{2} \nl
3848.21 & 3$d^2$D  $-$  5$p^2$P$^0$ & 8.86 & -1.56\tablenotemark{*} & fwC & 
	\nodata & 33 \ \ \ \ \ 7.26 \ \ \ \ \ 7.31 \nl
4390.57 & 4$p^2$P$^0$ $-$  5$d^2$D  & 10.00 & -0.50\tablenotemark{*} & fwD & 
	\phs71 \ \ \ \ \ 7.31 \ \ \ \ \ 7.34  & 
	60 \ \ \ \ \ 7.20 \ \ \ \ \ 7.21 \nl
\enddata
\tablecomments{The magnesium NLTE corrections have been computed using
the model atom by Gigas 1988.  Reference for the atomic data are from 
Fuhr \& Wiese 1998 (with estimated accuracies noted in capitals).}
\tablenotetext{*}{These lines are a clean blend of two to three lines from 
the same multiplet.   Their gf-values have been added together since the
wavelengths are nearly identical}. 
\end{deluxetable}

\clearpage
\begin{deluxetable}{lrll}
\footnotesize
\tablecaption{Elemental Abundances $\pm\sigma$ (\# lines) $\pm$systematic  \label{abu}}
\tablewidth{0pt}
\tablehead{
\colhead{Elem} & \colhead{Solar} & 
\colhead{Star cc} & \colhead{Star m}  
} 
\startdata
\ion{O}{1}  NLTE\tablenotemark{a} & 8.83 & 8.44 $\pm$0.09 (\ 3) {\it $\pm$0.12} & 8.28 $\pm$0.17 (\ 3) {\it $\pm$0.09}  \nl
\ion{Mg}{1} NLTE\tablenotemark{a} & 7.58 & 7.30 $\pm$0.02 (\ 2) {\it $\pm$0.20}\tablenotemark{b} & 
    		7.23 $\pm$0.13 (\ 3) {\it $\pm$0.20}\tablenotemark{b} \nl
\ion{Mg}{2} NLTE\tablenotemark{a} & 7.58 & 7.34 $\pm$0.00 (\ 1) {\it $\pm$0.10} & 7.26 $\pm$0.06 (\ 2) {\it $\pm$0.06} \nl
\ion{Si}{2}      & 7.56 & 7.18 $\pm$0.19 (\ 2) {\it $\pm$0.14} & 7.03 $\pm$0.17 (\ 7) {\it $\pm$0.18}  \nl
\ion{Sc}{2}      & 3.10 & 2.99 $\pm$0.09 (\ 2) {\it $\pm$0.30} & 2.64 $\pm$0.00 (\ 1) {\it $\pm$0.22}  \nl 
\ion{Ti}{2}      & 4.94 & 4.55 $\pm$0.23 (13)  {\it $\pm$0.25} & 4.16 $\pm$0.16 (18)  {\it $\pm$0.15}  \nl
\ion{Cr}{2}      & 5.69 & 5.22 $\pm$0.13 (10)  {\it $\pm$0.14} & 5.17 $\pm$0.17 (10)  {\it $\pm$0.07} \nl
\ion{Fe}{1}      & 7.50 & 7.15 $\pm$0.30 (\ 4) {\it $\pm$0.32} & 7.03 $\pm$0.24 (\ 5) {\it $\pm$0.23}  \nl
\ion{Fe}{2}      & 7.50 & 7.10 $\pm$0.15 (20)  {\it $\pm$0.17} & 6.92 $\pm$0.16 (35)  {\it $\pm$0.13} \nl
\ion{Ni}{2}      & 6.25 & \nodata                              & 5.49 $\pm$0.06 (\ 2) {\it $\pm$0.05} \nl
\ion{Sr}{2}      & 2.93 & \nodata                              & 1.71 $\pm$0.00 (\ 1) {\it $\pm$0.25} \nl 
\enddata
\tablecomments{Two uncertainties are listed for each element per star.   First is the 
line-to-line scatter in the abundances ($\sigma$) followed by the number of lines in the
average.  Second is an estimate of the potential systematic errors due to uncertainties in
the atmospheric parameters (see Table~\ref{abu-unc}).   This second uncertainty is noted
in italics and it is probably an overestimate (adding in quadrature does not account for 
\teff-gravity pairs that can be discounted).}
\tablenotetext{a}{Oxygen NLTE corrections of $-$0.2 and $-$0.3 have
been applied to Star cc and Star m, respectively, as determined by
Przybilla \etal (2000).   Magnesium NLTE corrections are discussed
further in Table~\ref{mgnlte}.}   
\tablenotetext{b}{The systematic uncertainty for \ion{Mg}{1} has been set to
$\pm$0.20 after considering only the allowed \teff-gravity pairs.}
\end{deluxetable}


\clearpage
 
\begin{deluxetable}{lccc}
\footnotesize
\tablecaption{Abundance Uncertainties \label{abu-unc}}
\tablewidth{0pt}
\tablehead{
\colhead{Elem} & \colhead{$\Delta$T=+150\,K} & 
                 \colhead{$\Delta$log~g=$-$0.1} &  
                 \colhead{$\Delta$$\xi=-$1\,km\,s$^{-1}$} \\[.2ex]
\colhead{    } & \colhead{Star cc / Star m} & 
                 \colhead{Star cc / Star m} & 
                 \colhead{Star cc / Star m}  } 
\startdata
O  I  &  0.08 / 0.06  & 0.08 / 0.05  & 0.04 / 0.04 \nl
Mg I  &  0.27 / 0.22  & 0.19 / 0.16  & 0.07 / 0.01  \nl
Mg II &  0.03 / 0.03  & 0.07 / 0.05  & 0.06 / 0.02  \nl
Si II & $-$0.03 / $-$0.02 & \phs0.02 / $-$0.01 & 0.14 / 0.16 \nl
Sc II &  0.24 / 0.20  & 0.13 / 0.08  & 0.13 / 0.01  \nl
Ti II &  0.18 / 0.14  & 0.10 / 0.05  & 0.14 / 0.03 \nl
Cr II &  0.09 / 0.06  & 0.05 / 0.01  & 0.09 / 0.04 \nl
Fe I  &  0.26 / 0.20  & 0.18 / 0.11  & 0.04 / 0.01  \nl
Fe II &  0.08 / 0.05  & 0.04 / 0.01  & 0.15 / 0.12  \nl
Ni II &  \nodata / 0.01 & \ \nodata / $-$0.02 & \nodata / 0.04  \nl
Sr II &  \nodata / 0.23 &  \nodata / 0.10 & \nodata / 0.01  \nl
\enddata
\end{deluxetable}

\clearpage
\begin{deluxetable}{lllrrccc}
\footnotesize
\tablecaption{Differential Abundances \label{diff}}
\tablewidth{0pt}
\tablehead{
\colhead{Elem \ \ \ \ \ \ \ Solar } & 
\colhead{Gal AIs\tablenotemark{*}} &
\colhead{SMC\tablenotemark{*}} &
\colhead{Star cc} & \colhead{Star m} &
\colhead{Mean(cc,m)} & \colhead{Mean(cc,m)} & 
\colhead{Mean(cc,m)} \\[.2ex]
\colhead{ } & 
\colhead{$-$Solar} & \colhead{$-$Solar} & 
\colhead{$-$Solar} & \colhead{$-$Solar} & 
\colhead{$-$Solar} & \colhead{$-$GAL AIs} & 
\colhead{$-$SMC AIs} 
} 
\startdata
\ion{O}{1}  \ \ \ \ \ \ \ \ \ \ 8.83 & $-$0.24 & $-$0.69  &  $-$0.39 & $-$0.55 & $-$0.47 & $-$0.23 & +0.22 \nl
\ion{Mg}{1} \ \ \ \ \ \ \ \     7.58 & $-$0.10 & $-$0.76  &  $-$0.28 & $-$0.35 & $-$0.32 & $-$0.22 & +0.45 \nl
\ion{Mg}{2} \ \ \ \ \ \ \       7.58 & $-$0.12 & $-$0.75  &  $-$0.24 & $-$0.32 & $-$0.28 & $-$0.16 & +0.47 \nl
\ion{Si}{2} \, \ \ \ \ \ \ \    7.56 & $-$0.22 & $-$0.58  &  $-$0.38 & $-$0.53 & $-$0.46 & $-$0.24 & +0.13 \nl
\ion{Sc}{2} \ \ \ \ \ \ \ \     3.10 &   +0.04 & $-$0.81  &  $-$0.11 & $-$0.46 & $-$0.29 & $-$0.33 & +0.52 \nl 
\ion{Ti}{2} \ \ \ \ \ \ \ \     4.94 & $-$0.07 & $-$0.64  &  $-$0.39 & $-$0.78 & $-$0.59 & $-$0.52 & +0.08 \nl
\ion{Cr}{2} \ \ \ \ \ \ \ \     5.69 & $-$0.07 & $-$0.71  &  $-$0.47 & $-$0.52 & $-$0.50 & $-$0.43 & +0.21 \nl
\ion{Fe}{1} \ \ \ \ \ \ \ \ \   7.50 &   +0.05 & $-$0.89  &  $-$0.35 & $-$0.47 & $-$0.41 & $-$0.46 & +0.48 \nl
\ion{Fe}{2} \ \ \ \ \ \ \ \     7.50 & $-$0.11 & $-$0.73  &  $-$0.40 & $-$0.58 & $-$0.49 & $-$0.38 & +0.24 \nl
\ion{Ni}{2} \ \ \ \ \ \ \ \  6.25 & \ \nodata & \ \nodata & \ \nodata & $-$0.76 & $-$0.76\tablenotemark{a} & \ \nodata & \ \nodata \nl
\ion{Sr}{2} \ \ \ \ \ \ \ \     2.93 & $-$0.52 & $-$1.72  & \ \nodata & $-$1.22 & $-$1.22\tablenotemark{a} & $-$0.70\tablenotemark{a} & +0.50\tablenotemark{a} \nl 
\nl
\ion{O}{1}/\ion{Fe}{2}      & $-$0.13 & +0.04   & +0.01     & +0.03 & +0.02 & +0.15 & $-$0.02 \nl
$\alpha$(Mg,Si)/\ion{Fe}{2} & $-$0.04 & +0.03   & +0.09     & +0.16 & +0.13 & +0.17 &   +0.10 \nl
\ion{Sr}{2}/\ion{Fe}{2}     & $-$0.41 & $-$0.99 & \ \nodata & $-$0.64 & $-$0.73\tablenotemark{a} & $-$0.32\tablenotemark{a} & +0.26\tablenotemark{a} \nl
\enddata
\tablenotetext{*}{The Galactic and SMC A-supergiant 
  abundances are those discussed in Table~6 of Venn 1999.}
\tablenotetext{a}{``Mean'' value is from only one star. }
\end{deluxetable}

\clearpage
\begin{deluxetable}{lcr}
\footnotesize
\tablecaption{Distances and Oxygen Abundances \label{dist}}
\tablewidth{0pt}
\tablehead{
\colhead{Object} & \colhead{12+log(O/H)} & \colhead{D (kpc)} 
} 
\startdata
STARS \nl
Star m  & 8.28 $\pm$0.17 & 0.28 \nl
Star cc & 8.44 $\pm$0.09 & 0.27 \nl
\nl
\ion{H}{2} \nl
Ho 11  & 8.24 $\pm$0.20 & 0.78 \nl
Ho 12  & 8.23 $\pm$0.20 & 0.85 \nl
Ho 3*   & 8.21 $\pm$0.15 & 0.77 \nl
Ho 6/8* & 8.20 $\pm$0.09 & 0.85 \nl
Ho 14  & 8.27 $\pm$0.20 & 1.28 \nl
Ho 13  & 8.50 $\pm$0.20 & 1.45 \nl
Ho 15*  & 8.11 $\pm$0.12 & 1.59 \nl
Nucleus & 8.44 $\pm$0.25 & 0.06 \nl
\nl
PN \nl
S33    & 8.01 $\pm$0.16 & 0.30 \nl
S16    & 8.10 $\pm$0.08 & 0.49 \nl
\enddata
\tablecomments{Oxygen abundances for the \ion{H}{2} 
regions from Pagel \etal (1980; a * notes those with 
\ion{O}{3} measurements), and for the 
planetary nebulae from Richer \& McCall (1995).
Distances are calculated using NGC\,6822 
\ion{H}{1} dynamical center coordinates 
(19$^h$ 42$^m$ 06.7$^s$, $-$14$^o$ 55' 22.0'', 1950), 
position angle (122$^o$), and inclination (50.1$^o$)
from Brandenburg \& Skillman (1998).}
\end{deluxetable}


\clearpage
\clearpage

\clearpage
\clearpage

\clearpage
\clearpage

\clearpage

\clearpage
\clearpage

\clearpage

\clearpage
\clearpage

\clearpage

\end{document}